\documentclass{article}
\usepackage{setspace}
\usepackage[utf8]{inputenc}
\usepackage{amsfonts,amsmath,amssymb,mathtools}
\usepackage[left=2cm,right=2cm,top=2cm,bottom=2cm,bindingoffset=0cm]{geometry}
\usepackage{graphicx}
\usepackage{authblk}
\usepackage{xcolor}
\usepackage{hyperref}
\usepackage{authblk}
\usepackage{slashed}
\usepackage{braket}
\usepackage{comment}
\usepackage{pb-diagram}
\usepackage{wasysym}
\usepackage{amsthm}
\usepackage{cancel}
\usepackage{tikz}\usetikzlibrary{tikzmark}
\usetikzlibrary{fadings} 
\usetikzlibrary{positioning}
\usetikzlibrary{backgrounds} 
\usetikzlibrary{decorations.pathreplacing}
\usetikzlibrary{arrows}
\definecolor{airforceblue}{rgb}{0.36, 0.54, 0.66}	
\definecolor{beige}{rgb}{0.96, 0.96, 0.86}
\definecolor{bittersweet}{rgb}{1.0, 0.44, 0.37}
\definecolor{melon}{rgb}{0.99, 0.74, 0.71}
\definecolor{mustard}{rgb}{1.0, 0.86, 0.35}
\definecolor{lava}{rgb}{0.81, 0.06, 0.13}
\definecolor{magnolia}{rgb}{0.97, 0.96, 1.0}
\definecolor{lavendermist}{rgb}{0.9, 0.9, 0.98}
\definecolor{lavendergray}{rgb}{0.77, 0.76, 0.82}
\definecolor{palepink}{rgb}{0.98, 0.85, 0.87}
\definecolor{palesilver}{rgb}{0.79, 0.75, 0.73}
\definecolor{cadetgrey}{rgb}{0.57, 0.64, 0.69}
\definecolor{anti-flashwhite}{rgb}{0.95, 0.95, 0.96}
\colorlet{Light0anti-flashwhite}{anti-flashwhite!70!white}
\colorlet{Lightanti-flashwhite}{anti-flashwhite!50!white}
\colorlet{Light2anti-flashwhite}{anti-flashwhite!30!white}
\definecolor{linkcolor}{rgb}{0,0,1}
\definecolor{urlcolor}{rgb}{0,0,1}

\usepackage{tikz-cd}
\usetikzlibrary{decorations.markings}
\usetikzlibrary{positioning}
\usepackage{braids}
\usepackage{array}
\usepackage{ytableau}
\usepackage{longtable}
\usepackage{empheq}
\usepackage{multicol}
\usepackage{mathrsfs}
\usepackage{diagbox}
\usepackage[vcentermath]{youngtab}
\usepackage[natbib=true, backend = bibtex, style=numeric-comp, sorting=none]{biblatex}
\hypersetup{pdfstartview=FitH,  linkcolor=linkcolor,urlcolor=urlcolor,citecolor=urlcolor, colorlinks=true}

\newcommand\bem{\begin{pmatrix}}
\newcommand\eem{\end{pmatrix}}
\newcommand\beq{\begin{equation}}
\newcommand\eeq{\end{equation}}
\newcommand\beqs{\begin{equation*}}
\newcommand\eeqs{\end{equation*}}

\newcommand{\tr}{\text{tr}}

\date{}
\begin{document}

\title{\bf Implications for colored HOMFLY polynomials from explicit formulas for group-theoretical structure }
\author[1,2]{{\bf E.~Lanina}\thanks{\href{mailto:lanina.en@phystech.edu}{lanina.en@phystech.edu}}}
\author[1,2,3]{{\bf A.~Sleptsov}\thanks{\href{mailto:sleptsov@itep.ru}{sleptsov@itep.ru}}}
\author[1,2]{{\bf N.~Tselousov}\thanks{\href{mailto:tselousov.ns@phystech.edu}{tselousov.ns@phystech.edu}}}

\vspace{5cm}

\affil[1]{Moscow Institute of Physics and Technology, 141700, Dolgoprudny, Russia}
\affil[2]{Institute for Theoretical and Experimental Physics, 117218, Moscow, Russia}
\affil[3]{Institute for Information Transmission Problems, 127994, Moscow, Russia}
\renewcommand\Affilfont{\itshape\small}

\maketitle

\vspace{-7cm}

\begin{center}
	\hfill ITEP-TH-31/21\\
	\hfill IITP-TH-20/21\\
	\hfill MIPT-TH-17/21
\end{center}

\vspace{4cm}

\begin{abstract}


We have recently proposed \cite{LST1} a powerful method for computing group factors of the perturbative series expansion of the Wilson loop in the Chern-Simons theory with $SU(N)$ gauge group. In this paper, we apply the developed method to obtain and study various properties, including nonperturbative ones, of such vacuum expectation values. 

First, we discuss the computation of Vassiliev invariants. Second, we discuss the Vogel theorem of not distinguishing chord diagrams by weight systems coming from semisimple Lie (super)algebras. Third, we provide a method for constructing linear recursive relations for the colored Jones polynomials considering a special case of torus knots $T[2,2k+1]$. Fourth, we give a generalization of the one-hook scaling property for the colored Alexander polynomials. And finally, for the group factors we provide a combinatorial description, which has a clear dependence on the rank $N$ and the representation $R$. 
\end{abstract}

\setcounter{equation}{0}
\section{Introduction}\label{intro}
Knots and links arise in various areas of modern theoretical and mathematical physics \cite{Witten1988hf, Chern1974ft, Guadagnini:1989kr, Guadagnini1989ma, Kaul1991np, RamaDevi1992np, Ramadevi1993np, Ramadevi1994zb, TURAEV1992865, Mironov2011aa, Anokhina2012ae, Anokhina2013ica, Zodinmawia2011oya, Zodinmawia2012kn, Ooguri1999bv, Labastida1997uw, Labastida2000yw, Labastida2001ts, Marino2001re, Alexander, Conway, Jones}.  A way to research the knotted/linked structures is the study of knot/link invariants. Due to their connections to physical and mathematical theories, the knot invariants have different interpretations and definitions in terms of the theories. These connections present an intriguing phenomenon that can be particularly useful in research contexts; namely, a progress in study of the knot invariants sometimes can be reformulated as a new nontrivial result in other connected areas. 

In this paper, we discuss properties of a family of quantum knot invariants, the so-called colored HOMFLY polynomials. Our analysis is carried out in terms of the perturbative expansion \cite{Guadagnini:1989kr, Guadagnini1989ma, GUADAGNINI1990575, Labastida1997uw}:
\begin{equation}
\label{Pert expan}
     \mathcal{H}_{R}^{\mathcal{K}}\left(q, a\right)\Big|_{q = e^{\hbar},\, a = e^{N \hbar}} = \sum_{n=0}^{\infty} \left( \sum_{m=1}^{\text{dim}\,\mathbb{G}_{n}} \mathcal{V}^{\mathcal{K}}_{n,m} \ \mathcal{G}_{n, m}^{R} \right) \hbar^{n}.
\end{equation}
Formulas of this kind for the colored HOMFLY polynomials are well-known in literature and can be understood in two different ways. First, as it was established in \cite{Witten}, the colored HOMFLY polynomials are vacuum expectation values of Wilson loop operators in the 3d Chern-Simons theory (see Section~\ref{CS theory}). Formula \eqref{Pert expan} follows directly as the perturbative expansion in the coupling constant $\hbar$ after an appropriate gauge fixing procedure. Second, formula \eqref{Pert expan} can be viewed as a more general universal Vassiliev knot invariant/the Kontsevich integral \cite{Kontsevich, Vassiliev, chmutov_duzhin_mostovoy_2012} computed in a special point (see Section~\ref{Kontsevich integral}). 

The colored HOMFLY polynomial $\mathcal{H}_{R}^{\mathcal{K}}\left(q, a\right)$ has two parameters: 1) a knot $\mathcal{K}$, and 2) an irreducible finite dimensional representation $R$ of $SU(N)$. The dependence on these two parameters {\it appears to split} in \eqref{Pert expan}. This is the main feature of the perturbative expansion. The knot dependent functions $\mathcal{V}^{\mathcal{K}}_{n,m}$ are called \textit{Vassiliev invariants} or \textit{finite type invariants} \cite{Vassiliev, chmutov_duzhin_mostovoy_2012}. The group dependent functions $\mathcal{G}^{R}_{n,m}$ are called {\it group factors}, and $\text{dim}\,\mathbb{G}_{n}$ is a number of linearly independent group factors $\mathcal{G}_{n, m}^{R}$ at a certain level $n$. 

While the structure of Vassiliev invariants remains mysterious, in previous work~\cite{LST1} we have closely approached a complete description of the HOMFLY group structure and provided an \textit{algorithm to compute the group factors} that are valid for {\it any} representation $R$ and an {\it arbitrary} value of $N$. \\

This paper is devoted to applications of the new knowledge on group factors to the research of {\it new properties} of quantum knot invariants.  Below we present an incomplete list of main possible applications that were considered in the present work:
\begin{itemize}
    \item Vassiliev invariants are practically computed via perturbative expansions of the colored HOMFLY or Kauffman polynomials. To perform such computations one uses two ingredients. First, many knot polynomials in different (large enough) representations are known. Second, an explicit form of the corresponding group factors has been obtained~\cite{LST1}. An explicit form of the HOMFLY group factors was available only for $\mathfrak{sl}_N$ fundamental representation and any $\mathfrak{sl}_2$ representation \cite{chmutov_duzhin_mostovoy_2012}, so that Vassiliev invariants were obtained only up to the 6-th order~\cite{katlas}. Our result~\cite{LST1} allows one to compute Vassiliev invariants \textit{of higher orders} (see Section~\ref{MoreVI}), which is crucial in some cases; for example, mutant knots \cite{Morton, Bishler2020fcy, Bishler2020wbq} have the same Vassiliev invariants up to the 10-th order.

    \item The second important question which we consider is the problem of \textit{not distinguishing Vassiliev invariants} by the weight systems of semisimple Lie (super)algebras (see Section~\ref{Vogel}), that was brought up in Vogel's work~\cite{VOGEL20111292}. It is believed that a set of all Vassiliev invariants is a complete knot invariant; that is, one can distinguish any two knots knowing the sets of Vassiliev invariants for them.
    
    All Vassiliev invariants are included into the Kontsevich integral, which is a series with an infinite number of chord diagrams. A basis in the space of chord diagrams must be found in each order separately. So, it is not convenient to deal with the Kontsevich integral. Thus, one usually works not with the Kontsevich integral (see Section~\ref{Kontsevich integral}), but with its image after applying the Lie algebras' weight system. For example, it is convenient because of the fact that the $\mathfrak{sl}_N$ weight system applied to the Kontsevich integral is the HOMFLY polynomial, and the $\mathfrak{so}_N$ weight system gives the so-called Kauffman polynomial. These facts allow us to get the corresponding Vassiliev invariants explicitly, as lots of methods for calculating these polynomials were developed, including modern Reshetikhin-Turaev approach, evolution method, and differential expansion \cite{Alekseev:2019nzw, Alekseev:2019klg, Morozov:2018fnb, Bai:2018twf, Dunin-Barkowski:2017zsd, Bishler:2017ifx, Dhara:2017ukv, Bai:2017nqe, Mironov:2017dnr, Mironov:2016pyz, Bishler:2020kqw, Morozov:2020adq}. 
    
    However, as it was shown by P. Vogel~\cite{VOGEL20111292}, \textit{not all of the Vassiliev invariants can be obtained from the only HOMFLY polynomial}. Thus, we face with a problem of generalization of the HOMFLY group factors to all simple Lie algebras by the use of Vogel's parametrization. The solution of this problem will have allowed us to answer the question, whether all of the Lie algebras' weight systems in whole distinguish all Vassiliev invariants.
    \item The third main question is a research on \textit{recursive} relations for the HOMFLY polynomials (for a previous work see~\cite{garoufalidis2018colored}) It is important because of the subsequent ability of finding the HOMFLY polynomials for higher representations knowing the HOMFLY polynomials for a set of smaller representations. In other words, it provides another method of calculating the HOMFLY polynomials for higher representations. However, it is a rather difficult problem, and in this paper we start its investigation with the case of the Jones polynomials (see Section~\ref{Jones}).
     \item The colored Alexander polynomial (which is the colored HOMFLY polynomial in a special point $a = 1$) has the so called \textit{1-hook scaling property} \cite{Mishnyakov:2019otq, MIRONOV2018268}:
    \begin{equation}\label{1-hook-SP}
    \mathcal{A}_{R}^{\mathcal{K}}(q)-\mathcal{A}_{[1]}^{\mathcal{K}}\left(q^{|R|}\right)=0, \quad \text { where } R=\left[r, 1^{L}\right].
    \end{equation}
    This symmetry is particularly interesting because it relates the Alexander polynomial for any 1-hook representation, which is non-rectangular in general, with the Alexander polynomial for the fundamental representation. Using our methods we have managed to \textit{generalize} this property to \textit{any} representation (see Subsection~\ref{AlScaling}).
    \item We present a new refined algorithm that allows one to compute the HOMFLY group factors in any representation $R$ and rank $N$ (see Subsection~\ref{CombConstr}). The new algorithm reduces a computation of a group factor to a simple combinatorial problem. It has two advantages compared with the previous method \cite{LST1}. First, the new algorithm provides a {\it unified} approach to all group factor. The method described in~\cite{LST1} was formulated for even and odd basis elements separately. There was no precise algorithm for eliminating negative powers of $N$ from the odd elements.  Second, the formula for the new algorithm \eqref{combcasimir} has clear dependence on the rank $N$ and the representation $R$.
\end{itemize}
\newpage
This paper is organized as follows. In Section \ref{colored HOMFLY}, we provide basic facts and definitions on the colored HOMFLY polynomials and the group factors. The beginning of Section \ref{HOMFLY-GF} and Subsection \ref{Z-GF} are devoted to a brief introduction into the algorithm for computing group factors in any representation $R$. In Subsection \ref{CombConstr}, we present a new combinatorial method for another group factors' basis. Also, in Subsection \ref{GF-Jacobi}, we describe a basis in group factors corresponding to the Jacobi diagrams. The next Sections \ref{Vogel} and \ref{MoreVI} are devoted to distinguishing and computing Vassiliev invariants using the last basis in group factors. In Section \ref{Jones}, we develop a new method of calculation of recursive relations for the Jones polynomials and apply it to the special case of torus knots $T[2,2k+1]$. In Section \ref{Alexander}, we present consequences for another HOMFLY speciality -- the Alexander polynomial. Specifically, in Subsection~\ref{NewSymmAlexander} we notice an unknown symmetry of the Alexander polynomial which follows from the conjugation symmetry of the HOMFLY polynomial. Then, we discuss that the Alexander polynomial can be deformed into the HOMFLY polynomial in Subsection~\ref{AlexanderDef}. And, finally, in Subsection~\ref{AlScaling} we provide a new interesting property of the colored Alexander polynomial that generalizes 1-hook scaling property. We conclude our analysis in Section \ref{discussion} by a discussion of the results and possible future research directions.

\setcounter{equation}{0}
\section{Colored HOMFLY polynomial}\label{colored HOMFLY}
Colored HOMFLY polynomials are topological invariants of knots and links. They generalize many known polynomial knot invariants, such as well-known Jones and Alexander polynomials, quantum Reshetikhin-Turaev invariants of $\mathfrak{sl}_N$. An essence of this generalization is a clever analytical continuation and the introduction of a new variable $a = q^N$. This analytical continuation allows one to connect the colored HOMFLY polynomials with quantum knot invariants not only of Lie groups, but also of Lie supergroups \cite{queffelec2015note}.

In the subsections below, we provide \textit{two different self-consistent definitions of the colored HOMFLY polynomial}.
\begin{itemize}
    \item The first one is \textit{physical}, as it provides the connection with the 3d Chern-Simons theory.
    \item The second one is \textit{mathematical} and allows one to obtain precise results and prove theorems in the theory of quantum knot invariants.
\end{itemize}
It is simpler to keep in mind the physical definition, but in order to provide explicit statements, we utilize the mathematical definition.

\subsection{Chern-Simons theory} \label{CS theory}
The \textit{colored HOMFLY polynomial} is the vacuum expectation value of the Wilson loop operator in the 3d Chern-Simons theory with the gauge group $SU(N)$:
     \begin{equation}
     \label{WilsonLoopExpValue}
         \mathcal{H}_{R}^{\mathcal{K}}(q, a) = \frac{1}{\text{qdim}(R)}\left\langle \text{tr}_{R} \ \text{Pexp} \left( \oint_{\mathcal{K}} A \right) \right\rangle_{\text{CS}},
     \end{equation}
     where Pexp denotes a path-ordered exponential and the Chern-Simons action is given by
     \begin{equation}\label{CSAction}
         S_{\text{CS}}[A] = \frac{\kappa}{4 \pi} \int_{S^3} \text{tr} \left(  A \wedge dA +  \frac{2}{3} A \wedge A \wedge A \right).
     \end{equation}
     The contour of the Wilson loop operator is a knot $\mathcal{K}$ and $R$ is an irreducible finite dimensional representation of the algebra $\mathfrak{sl}_N$, that corresponds to $SU(N)$ gauge group, $\text{qdim}(R)$ is a quantum dimension. In formulas ~\eqref{WilsonLoopExpValue} and~\eqref{CSAction} the gauge field has the form $A=A^kT_k$, where $T_k$ are $\mathfrak{sl}_N$ generators. Remarkably, the complicated correlator \eqref{WilsonLoopExpValue} is a polynomial knot invariant in two variables $q$ and $a$ that are parametrized as follows:
         $q = e^{\hbar}, \ a = e^{N \hbar}, \ \hbar := \frac{2 \pi i}{\kappa + N}\ $.
         
     The formula \eqref{Pert expan} naturally appears from \eqref{WilsonLoopExpValue} when the Wilson loop operator is expanded in the parameter $\hbar$:
     \begin{align}
     \begin{aligned}
     \label{LoopExpansionHOMFLY}
      \mathcal{H}_{R}^{\mathcal{K}} &= \sum\limits_{n=0}^{\infty} \oint dx_{1}\int
      dx_{2}...\int dx_{n} \langle\,A^{a_1}(x_{1})A^{a_2}(x_{2})...A^{a_n}(x_{n})\,\rangle\,
     \tr_R(T_{a_1} T_{a_2}...T_{a_n}) \, \hbar^n = \\
     & \phantom{\hspace{35mm}}= \boxed{\sum_{n=0}^{\infty} \left( \sum_{m=1}^{\text{dim}\,\mathbb{G}_{n}} \mathcal{V}^{\mathcal{K}}_{n,m} \ \mathcal{G}_{n, m}^{R} \right) \hbar^{n},}
     \end{aligned}
     \end{align}
     where $\text{dim}\,\mathbb{G}_{n}$ is a number of linearly independent group factors $\mathcal{G}_{n, m}^{R}$ at a certain level $n$. Knot dependent functions -- Vassiliev invariants $\mathcal{V}_{n,m}$, are contour integrals along the knot. The group factors are traces in the representation $R$ of several $\mathfrak{sl}_N$ generators $T_a$. The first nontrivial example of the group factor:
     \begin{equation}
     \label{G21}
         \mathcal{G}_{2, 1}^{R} = \text{tr}_R \left(\sum_{a,b}   T_a T_b T_a T_b  -  T_a T_a T_b T_b \right) = N \mathcal{C}_2^{R}\,,
     \end{equation}
     where the generators are normalized as $\text{tr}_R \left( T_a T_b \right) = \frac{\delta_{ab}}{2 \dim R} \,$, and $\mathcal{C}_2^{R}$ is an eigenvalue of the quadratic Casimir operator of $\mathfrak{sl}_N$. In this particular example one can observe a remarkable property of the group factors, namely, they are functions of the eigenvalues of the Casimir operators. The expressions under traces in group factors lie in the center of the universal enveloping algebra $ZU(\mathfrak{sl}_N)$ (for a proof see Ch.6.1.2 in~\cite{chmutov_duzhin_mostovoy_2012}). The basis in $ZU(\mathfrak{sl}_N)$ is multiplicatively generated by Casimir operators $\hat{\mathcal{C}}_k, \ k = 1, \ldots, N$. Hence being the traces and taken in a certain representation $R$, group factors are polynomials of their eigenvalues in the representation $R\,$:
     \begin{equation}\label{GPol}
         \boxed{\mathcal{G}_{n, m}^{R} = \text{Pol}\left( \mathcal{C}_1^R, \ldots, \mathcal{C}_{N}^R \right).}
     \end{equation}
     
     Multiplication of group factors respects the level, that is the sum of the factors' levels is a level of the resulting group factor. 
     For example, we can set the first group factor of the 4-th level to be the square of the group factor of the second level:
     \begin{equation}
     \mathcal{G}^R_{4,1}=\left(\mathcal{G}^R_{2,1}\right)^2.    
     \end{equation}
     
     Group factors that cannot be represented as products of other group factors are called \textit{primary group factors}. Vassiliev invariants corresponding to primary group factors are called \textit{primary} or \textit{primitive Vassiliev invariants}. 

\subsection{Kontsevich integral } \label{Kontsevich integral}
The \textit{Kontsevich integral} or the \textit{universal Vassiliev invariant} \cite{chmutov_duzhin_mostovoy_2012, Kontsevich} is a more general structure than the HOMFLY polynomial:
     \begin{equation}
         I^{\mathcal{K}} = \sum_{n = 0}^{\infty} \left( \sum_{m} \ \mathcal{V}_{n,m}^{\mathcal{K}} \ \mathcal{D}_{n,m} \right)\hbar^n,
     \end{equation}
     where the group factors $\mathcal{G}^{R}_{n,m}$ are generalized to \textit{chord diagrams} $\mathcal{D}_{n,m}$ with $n$ chords, while $\mathcal{V}_{n,m}^{\mathcal{K}}$ are Vassiliev invariants. At each level $n$ there are $\dim \mathcal{D}_n$ chord diagrams, hence $m = 1,\ldots, \dim \mathcal{D}_n$.
     We provide examples of chord diagrams of small degrees in Fig.~\ref{fig:ChordDiag}.
\begin{figure}[h!]
\begin{center}
     \begin{tikzpicture}
        \draw[very thick] circle (0.7cm);
    \end{tikzpicture}
    \hspace{10mm}
    \begin{tikzpicture}
        \foreach \x [count=\p] in {0,...,3} {
            \node[shape=circle,fill=black, scale=0.4] (\p) at (45-\x*90:0.7) {};
        };
        \draw[very thick] (1) arc (45:360 + 45:0.7);
        \draw[very thick] (1) to (3);
        \draw[very thick] (2) to (4);
    \end{tikzpicture}
    \hspace{10mm}
     \begin{tikzpicture}
        \foreach \x [count=\p] in {0,1,2,4,5,6} {
            \node[shape=circle,fill=black, scale=0.4] (\p) at (45-\x*45:0.7) {};
        };
        \draw[very thick] (1) arc (45:360 + 45:0.7);
        \draw[very thick] (1) to[bend right] (3);
        \draw[very thick] (4) to[bend right] (6);
        \draw[very thick] (5) to (2);
    \end{tikzpicture}
    \hspace{10mm}
     \begin{tikzpicture}
        \foreach \x [count=\p] in {0,...,11} {
            \node[shape=circle,fill=black, scale=0.4] (\p) at (45/2-\x*45:0.7) {};
        };
        \draw[very thick] (1) arc (45/2:360 + 45/2:0.7);
        \draw[very thick] (1) to[bend right] (3);
        \draw[very thick] (2) to[bend left] (8);
        \draw[very thick] (4) to[bend right] (6);
        \draw[very thick] (5) to[bend right] (7);
    \end{tikzpicture}
    \hspace{10mm}
    \begin{tikzpicture}
        \foreach \x [count=\p] in {0,...,11} {
            \node[shape=circle,fill=black, scale=0.4] (\p) at (45/2-\x*45:0.7) {};
        };
        \draw[very thick] (1) arc (45/2:360 + 45/2:0.7);
        \draw[very thick] (1) to[bend right] (3);
        \draw[very thick] (2) to[bend left] (7);
        \draw[very thick] (4) to[bend right] (6);
        \draw[very thick] (5) to[bend right] (8);
    \end{tikzpicture}
    \hspace{10mm}
     \begin{tikzpicture}
        \foreach \x [count=\p] in {0,...,11} {
            \node[shape=circle,fill=black, scale=0.4] (\p) at (45/2-\x*45:0.7) {};
        };
        \draw[very thick] (1) arc (45/2:360 + 45/2:0.7);
        \draw[very thick] (1) to (6);
        \draw[very thick] (2) to (5);
        \draw[very thick] (4) to (7);
        \draw[very thick] (3) to (8);
    \end{tikzpicture}
     \end{center}
\begin{equation}\nonumber
\label{chorddiag}
     \mathcal{D}_{0,1} \hspace{20mm} 
     \mathcal{D}_{2,1} \hspace{21mm} 
     \mathcal{D}_{3,1} \hspace{22mm} 
     \mathcal{D}_{4,1} \hspace{22mm} 
     \mathcal{D}_{4,2} \hspace{22mm} 
     \mathcal{D}_{4,3}
\end{equation}
    \caption{Examples of chord diagrams}
    \label{fig:ChordDiag}
\end{figure}
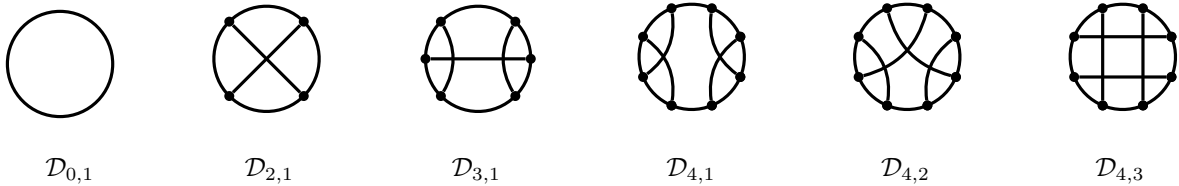\\
The space of chord diagrams $\mathcal{D}$ possesses a natural grading by the number of chords $\mathcal{D}=\bigoplus_{n}\mathcal{D}_n$. It is a well-known fact (see Ch.4 \cite{chmutov_duzhin_mostovoy_2012}) that the space of chord diagrams is an algebra with relations, the so-called 1T and 4T relations. The 4T relation provides a well-defined product in the space of chord diagrams, while the 1T relation forbids the chord diagrams with an isolated chord. In particular, there are no diagrams with one chord (see Fig.~\ref{fig:ChordDiag}).
One can return to group factors using a mapping called {\it Lie algebra weight system} $\varphi^{R}_{\mathfrak{sl}_N}$ associated with a representation $R\,$:  
     \begin{equation}\label{D-G}
         \varphi^{R}_{\mathfrak{sl}_N} \left( \mathcal{D}_{n,m} \right) = \mathcal{G}^{R}_{n,m}\,.
     \end{equation}
     As a corollary of this fact, one obtains the full colored HOMFLY polynomial as a special value of the Kontsevich integral:
     \begin{equation}\label{Z-H}
         \varphi_{\mathfrak{sl}_N}^{R}\left(I^{\mathcal{K}} \right) = \mathcal{H}_{R}^{\mathcal{K}}(q,a)\,.
     \end{equation}
     Chord diagrams give a unified approach to computing the group factors~\eqref{D-G}. Namely, one evaluates the unframed weight system $\varphi_{\mathfrak{sl}_N}^R$ associated with a representation $R$ on the basis elements $\mathcal{D}_{n,m}$ of the space of chord diagrams. The unframed condition for the weight system is needed due to the 1T relation in the algebra of chord diagrams. In other words, the unframed weight system sends a chord diagram with an isolated chord to zero.  We provide an example of the first non-trivial chord diagram $\mathcal{D}_{2,1}$ that is mapped to the group factor \eqref{G21}:
     \begin{equation}
         \varphi_{\mathfrak{sl}_N}^{R} \left( \mathcal{D}_{2,1} \right) = \text{tr}_R \left(\sum_{a,b}   T_a T_b T_a T_b  -  T_a T_a T_b T_b \right) = N \mathcal{C}_2^R\,.
     \end{equation}
     The second summand arises due to the deframing procedure \cite{chmutov_duzhin_mostovoy_2012}, while the first term is the image of the framed weight system. The framed weight system $\varphi^{R, \text{fr}}_{\mathfrak{sl}_N}$ can be computed for the chord diagram in a simple way that is clear from an example in Fig.~\ref{fig:FrSys}.
     \begin{figure}[h!]
     \begin{center}
        \begin{tikzpicture}[
        Circ/.style={draw,shape=circle,minimum size=14mm, node contents={},very thick}]
        \node (C2) [Circ];
        \draw[black, very thick]     (C2.north west) -- (C2.south east);
        \draw[black, very thick]     (C2.north east) -- (C2.south west);
        \fill[black]    (C2.north west) circle (2pt) node[above]{\bf{\scriptsize{a}}}
                        (C2.south west) circle (2pt) node[below]{\bf{\scriptsize{b}}}
                        (C2.south east) circle (2pt) node[below]{\bf{\scriptsize{a}}}
                        (C2.north east) circle (2pt) node[above]{\bf{\scriptsize{b}}};
        \end{tikzpicture}
     \end{center}
     \begin{equation}\nonumber
         \varphi^{R, \text{fr}}_{\mathfrak{sl}_N}(\mathcal{D}_{2,1}) = \text{tr}_{R} \left(\sum_{a,b} T_a T_b T_a T_b \right).
     \end{equation} 
      \caption{Example of applying the framed weight system}
         \label{fig:FrSys}
     \end{figure}
     
     The unframed weight system $ \varphi_{\mathfrak{sl}_N}^{R}$ associated with a representation $R$ can be represented as a sequence of maps:
     \begin{equation}
     \label{WeightSystem}
         \varphi_{\mathfrak{sl}_N}^{R} : \ \mathcal{D} \xrightarrow{\varphi_{\mathfrak{sl}_N}} Z U(\mathfrak{sl}_N) 
         \xrightarrow{\rho_R}  \text{End}(V_R) \xrightarrow{\text{Tr}_R}
         \mathbb{C}\,.
     \end{equation}
     At the first step, the unframed Lie algebra weight system $\varphi_{\mathfrak{sl}_N}$ maps a chord diagram $\mathcal{D}_{n,m}$ to the center of the universal enveloping algebra $Z U(\mathfrak{sl}_N)$. In other words, the image of a chord diagram is a combination of the Casimir operators $\hat{ \mathcal{C}}_k, \ k = 1, \ldots, N$. At the second step, one chooses all operators to be in a representation $R$. By the Schur lemma, the Casimir operators become proportional to the identity operator $\hat{\mathcal{C}}_k =  \mathcal{C}_k^{R} \, \hat{\text{I}}$. At the last step, one takes the trace in the representation $R$ and obtains a function in the eigenvalues of the Casimir operators:
     \begin{equation}
     \label{sln weight system}
     \boxed{
         \mathcal{G}_{n,m}^R = \varphi^{R}_{\mathfrak{sl}_N} \left( \mathcal{D}_{n,m} \right) = \text{Pol} \left( \mathcal{C}_1^R, \ldots, \mathcal{C}_{N}^R \right).
        }
     \end{equation}
      The Lie algebra weight system is a homomorphism, hence it respects the algebraic structure of the space of chord diagrams $\mathcal{D}$. However, it was shown that the Lie algebras weight systems have non-zero kernel \cite{VOGEL20111292}. This means that some Vassiliev invariants cannot be distinguished by the HOMFLY group factors. In this sense, chord diagrams are generalizations of group factors. However, in practical calculations chord diagrams are not useful because a basis in the space of chord diagrams is known explicitly only for few number of chords.
      
      It is convenient to introduce a filtration by the number of $\mathfrak{sl}_N$ generators in the center of the universal enveloping algebra $Z U(\mathfrak{sl}_N)$:
\begin{equation}
    \mathcal{Z}_2\subset \mathcal{Z}_3\subset \mathcal{Z}_4\subset\dots\subset Z U(\mathfrak{sl}_N)\, ,
\end{equation}
where $\mathcal{Z}_k$ consists of products of at most $2k$ generators. Then the mapping $\varphi_{\mathfrak{sl}_N}$~\eqref{D-G} sends $\mathcal{D}_n$ to $\mathcal{Z}_n$ and
\begin{equation}
    \varphi_{\mathfrak{sl}_N}^R: \ \mathcal{D}_n \xrightarrow{ \varphi_{\mathfrak{sl}_N}} \mathcal{Z}_n
         \xrightarrow{\rho_R}  \text{End}(V_R) \xrightarrow{\text{Tr}_R}
         \mathbb{C}\,.
\end{equation}
For a detailed description of the $\mathfrak{sl}_N$ weight system, in what follows we describe the embeddings $\varphi_{\mathfrak{sl}_N}^R(\mathcal{D}_n)\subset\mathcal{Z}_n$. 

\setcounter{equation}{0}
\section{HOMFLY group factors}\label{HOMFLY-GF}
It is tempting to describe the group structure of the colored HOMFLY polynomials explicitly and entirely. We would like to find such an amount of the HOMFLY polynomials' symmetries that is enough to fully determine each group factor in the perturbative expansion~\eqref{LoopExpansionHOMFLY}.

For a start, let us define the main tools that we use in our study of the HOMFLY group structure.
\begin{itemize}
    \item First, the colored HOMFLY polynomial has symmetries with respect to a representation $R$ and the $\hbar$ parameter. They restrict the functions $\mathcal{G}_{n,m}^R = \text{Pol} \left( \mathcal{C}_1^R, \ldots, \mathcal{C}_{N}^R \right)$~\eqref{sln weight system}.
    \item Second, we utilize the embedding $Z U(\mathfrak{sl}_N)\subset Z U(\mathfrak{gl}_\infty)$. In other words, we express $\mathfrak{sl}_N$ Casimir eigenvalues $\mathcal{C}_k^R$ in terms of $\mathfrak{gl}_\infty$ Casimir eigenvalues $C_k^R$.
\end{itemize}
In what follows, for $C_k^R$ we use two self-consistent formulas:
\begin{equation}
    \label{gl Casimirs}
    C_{k}^{R} = \sum_{i = 1}^{l(R)} \left( R_i - i + 1/2 \right)^k - \left(-i + 1/2 \right)^k,
\end{equation}
where $R_i$ are the entries of a Young diagram $R=[R_1,R_2,\dots,R_{l(R)}]$, and 
\begin{equation}\label{TTHC}
C_{k}^{R}=\sum_{i=1}^N \left(R_{i}-i+1/2\right)^{k}-\left(-i+1/2\right)^{k} +\sum\limits_{i=N+1}^{N+M} \left(\alpha_i-N-1/2\right)^k-\left(-\beta_i-N+1/2\right)^k,
\end{equation}
where for the representation $R$ we use the shifted Frobenius notation
\begin{equation}\label{TugTheHookYoungTNot}
\left[R_{1}, \ldots, R_{N}\right]\left(\alpha_{N+1}, \ldots, \alpha_{N+M} \mid \beta_{N+1}, \ldots, \beta_{N+M}\right).
\end{equation}
In the last expression $R_{i}$, $1 \leq i \leq N$, are lengths of the first $N$ rows of the Young diagram $R$ and the rest of the diagram is parametrized by shifted Frobenius variables:
\begin{equation}
\alpha_{i}=R_{i}-(i-N)+1, \quad
\beta_{i}=R_{i-N}^{T}-i+1, \quad i>N.
\end{equation}

\subsection{Algorithm for group factors}\label{Z-GF}
In this subsection, we briefly state the main results on the HOMFLY group structure obtained in~\cite{LST1}. We have constructed an algorithm of determining group factors in the loop expansion: 
\begin{equation}\label{HLoopExp}
    \mathcal{H}^{\mathcal{K}}_{R}=\sum\limits_{n=0}^{\infty}\hbar^n\sum\limits_{|\Delta|\leq n}\mathcal{C}_{\Delta}^{R}\sum_{m=0}^{n-|\Delta|} \left(v^{\mathcal{K}}_{\Delta,m}\right)_n N^m\,.
\end{equation}
The known HOMFLY symmetries (the tug-the-hook symmetry, the conjugation symmetry, the rank-level duality and the genus order restriction) dictate the following formulas for \textit{even elements}
\begin{equation}\label{AnalyticZeven}
    \mathcal{C}^R_{[2n]}:=\sum\limits_{k=1}^{2n}(-1)^k\frac{C^R_k \left(C^R_{2n-k}+2\theta^N_{2n-k}\right)}{2\cdot k!(2n-k)!}\,,
\end{equation}
where we define $\theta_k^N := \sum_{i = 1}^N \left( - i + 1/2 \right)^k$, and for \textit{odd elements}:
\begin{equation}\label{AnalyticZ}
    \mathcal{C}^R_{[2n+1]}=\sum\limits_{m=0}^{2n+1}\frac{(-1)^{m+1}}{N^m}\binom{2n+1}{m}  
    \left(
    \left(C^R_{2n-m+1}+\theta_{2n-m+1}^N \right) \left(C^R_1+\theta_1^N \right)^m-\theta_{2n-m+1}^N\left(\theta_1^N\right)^m 
    \right)\,.
\end{equation}
\begin{table}[h!]\caption{HOMFLY group factors up to the 9-th order}\label{GroupFactors}
\begin{center}
\renewcommand{\arraystretch}{2}
    \begin{tabular}{| m{1em} | m{21.5em} | m{1em} | m{21.5em} |} 
        \hline
        $\hbar^2$ & $\mathcal{C}_{[2]}$ & $\hbar^3$ &  $N\mathcal{C}_{[2]}$\\
        \hline 
        $\hbar^4$ & $\cancel{\boxed{\mathcal{C}_{[2]}}}$, $N^2\mathcal{C}_{[2]}$, $\mathcal{C}_{[2]}^2$, $\mathcal{C}_{[4]}$ 
        & $\hbar^5$ & $N\mathcal{C}_{[2]}$, $N^3\mathcal{C}_{[2]}$, $N\mathcal{C}^2_{[2]}$, $N\mathcal{C}_{[4]}$ \\
        \hline
        $\hbar^6$ &$\cancel{\boxed{\mathcal{C}_{[2]}}}$, $N^2\mathcal{C}_{[2]}$, $N^4\mathcal{C}_{[2]}$, $\cancel{\boxed{\mathcal{C}_{[2]}^2}}$, $N^2\mathcal{C}_{[2]}^2$, $\mathcal{C}_{[2]}^2 + 6\mathcal{C}_{[4]}$, $N^2\mathcal{C}_{[4]}$, $\mathcal{C}_{[2]}^3$, $\mathcal{C}_{[2]}\mathcal{C}_{[4]}$, $\mathcal{C}_{[6]}$ 
        &$\hbar^7$ & $N\mathcal{C}_{[2]}$, $N^3\mathcal{C}_{[2]}$, $N^5\mathcal{C}_{[2]}$, $N\mathcal{C}_{[2]}^2$ $N^3\mathcal{C}_{[2]}^2$, $ N\mathcal{C}_{[4]}$, $N^3\mathcal{C}_{[4]}$, $N\mathcal{C}_{[2]}^3$, $N\mathcal{C}_{[2]}\mathcal{C}_{[4]}$, $ \mathcal{C}_{[3,3]}$, $N\mathcal{C}_{[6]}$ \\
        \hline 
        $\hbar^8$ &$\cancel{\boxed{\mathcal{C}_{[2]}}}$, $N^2\mathcal{C}_{[2]}$, $N^4\mathcal{C}_{[2]}$, $N^6\mathcal{C}_{[2]}$, $\cancel{\boxed{\mathcal{C}_{[2]}^2}}$, $N^2\mathcal{C}_{[2]}^2$, $N^4\mathcal{C}_{[2]}^2$, $\mathcal{C}_{[2]}^2+6\mathcal{C}_{[4]}$, $N^2\mathcal{C}_{[4]},\; N^4\mathcal{C}_{[4]}$, $\cancel{\boxed{\mathcal{C}_{[2]}^3}}$, $N^2\mathcal{C}_{[2]}^3$, $\mathcal{C}_{[2]}(\mathcal{C}_{[2]}^2+6\mathcal{C}_{[4]})$, $N^2\mathcal{C}_{[2]}\mathcal{C}_{[4]}$, $ \mathcal{C}_{[2]}^3-90\,\mathcal{C}_{[6]}$,
        $N^2\mathcal{C}_{[6]}$, $N\mathcal{C}_{[3,3]}$, $\mathcal{C}_{[2]}^4$, $\mathcal{C}_{[2]}^2\mathcal{C}_{[4]}$, $\mathcal{C}_{[4]}^2$, $\mathcal{C}_{[2]}\mathcal{C}_{[6]}$, $\mathcal{C}_{[8]}$ 
        & $\hbar^9$ & $N\mathcal{C}_{[2]}$, $N^3\mathcal{C}_{[2]}$, $N^5\mathcal{C}_{[2]}$, $N^7\mathcal{C}_{[2]}$, $N\mathcal{C}_{[2]}^2$, $N^3\mathcal{C}_{[2]}^2$, $N^5\mathcal{C}_{[2]}^2$, $N\mathcal{C}_{[4]}$, $N^3\mathcal{C}_{[4]}$, $N^5\mathcal{C}_{[4]}$, $N\mathcal{C}_{[2]}^3$, $N^3\mathcal{C}_{[2]}^3$, $N\mathcal{C}_{[2]}\mathcal{C}_{[4]}$, $N^3\mathcal{C}_{[2]}\mathcal{C}_{[4]}$, $N\mathcal{C}_{[6]}$, $N^3\mathcal{C}_{[6]}$, $\mathcal{C}_{[3,3]}$, $N^2\mathcal{C}_{[3,3]}$, $N\mathcal{C}_{[2]}^4$, $N\mathcal{C}_{[2]}^2\mathcal{C}_{[4]}$, $N\mathcal{C}_{[4]}^2$, $\mathcal{C}_{[2]}\mathcal{C}_{[3,3]}$, $N\mathcal{C}_{[2]}\mathcal{C}_{[6]}$, $N\mathcal{C}_{[8]}$, $\mathcal{C}_{[5,3]}$ \\
        \hline
    \end{tabular}
\renewcommand{\arraystretch}{1}
\end{center}
\end{table}
\begin{table}[h!]\caption{primary HOMFLY group factors up to the 11-th order}\label{PrimGroupFactors}
\begin{center}
\renewcommand{\arraystretch}{2}
    \begin{tabular}{| m{1em} | m{21.5em} | m{1em} | m{21.5em} |} 
        \hline
        $\hbar^2$ & $\mathcal{C}_{[2]}$ & $\hbar^3$ &  $N\mathcal{C}_{[2]}$\\
        \hline 
        $\hbar^4$ &
        $N^2\mathcal{C}_{[2]}$, $\mathcal{C}_{[4]}$ 
        & $\hbar^5$ & $N\mathcal{C}_{[2]}$, $N^3\mathcal{C}_{[2]}$, $N\mathcal{C}_{[4]}$ \\
        \hline
        $\hbar^6$ &
        $N^2\mathcal{C}_{[2]}$, $N^4\mathcal{C}_{[2]}$,
        $\frac{1}{6}\mathcal{C}_{[2]}^2 + \mathcal{C}_{[4]}$, $N^2\mathcal{C}_{[4]}$, $\mathcal{C}_{[6]}$ 
        &$\hbar^7$ & $N\mathcal{C}_{[2]}$, $N^3\mathcal{C}_{[2]}$, $N^5\mathcal{C}_{[2]}$, $N\mathcal{C}^2_{[2]}$, $ N\mathcal{C}_{[4]}$, $N^3\mathcal{C}_{[4]}$, $ \mathcal{C}_{[3,3]}$, $N\mathcal{C}_{[6]}$  \\
        \hline 
        $\hbar^8$ &
        $N^2\mathcal{C}_{[2]}$, $N^4\mathcal{C}_{[2]}$, $N^6\mathcal{C}_{[2]}$,
        $N^2\mathcal{C}_{[2]}^2$, $\frac{1}{6}\mathcal{C}_{[2]}^2+\mathcal{C}_{[4]}$, $N^2\mathcal{C}_{[4]},\; N^4\mathcal{C}_{[4]}$,
         $\mathcal{C}_{[6]}+\frac{1}{15}\mathcal{C}_{[2]}\mathcal{C}_{[4]}$,
        $N^2\mathcal{C}_{[6]}$, $N\mathcal{C}_{[3,3]}$, $\mathcal{C}_{[8]}$ 
        & $\hbar^9$ & $N\mathcal{C}_{[2]}$, $N^3\mathcal{C}_{[2]}$, $N^5\mathcal{C}_{[2]}$, $N^7\mathcal{C}_{[2]}$, $N\mathcal{C}_{[2]}^2$, $N^3\mathcal{C}_{[2]}^2$, $N\mathcal{C}_{[4]}$, $N^3\mathcal{C}_{[4]}$, $N^5\mathcal{C}_{[4]}$, $N\mathcal{C}_{[2]}\mathcal{C}_{[4]}$, $N\mathcal{C}_{[6]}$, $N^3\mathcal{C}_{[6]}$, $\mathcal{C}_{[3,3]}$, $N^2\mathcal{C}_{[3,3]}$, $N\mathcal{C}_{[8]}$, $\mathcal{C}_{[5,3]}$ \\
        \hline
        $\hbar^{10}$ & $N^2\mathcal{C}_{[2]}$, $N^4\mathcal{C}_{[2]}$, $N^6\mathcal{C}_{[2]}$, $N^8\mathcal{C}_{[2]}$,\; $\frac{1}{6}\mathcal{C}_{[2]}^2+\mathcal{C}_{[4]}$, $N^2\mathcal{C}_{[2]}^2$, $N^4\mathcal{C}_{[2]}^2$,\; $N^2\mathcal{C}_{[4]}$, $N^4\mathcal{C}_{[4]}$, $N^6\mathcal{C}_{[4]}$,\;  $\mathcal{C}_{[6]}+\frac{1}{15}\mathcal{C}_{[2]}\mathcal{C}_{[4]}$, $N^2\mathcal{C}_{[2]}\mathcal{C}_{[4]}$,\; $N\mathcal{C}_{[3,3]}$, $N^3\mathcal{C}_{[3,3]}$,\; $\mathcal{C}_{[6]}-\frac{1}{90}\mathcal{C}^3_{[2]}$, $N^2\mathcal{C}_{[6]}$, $N^4\mathcal{C}_{[6]}$,\; $\mathcal{C}_{[8]}+\frac{1}{70}\mathcal{C}_4^2$, $N^2\mathcal{C}_{[8]}$,\;$\mathcal{C}_{[8]}+\frac{1}{28}\mathcal{C}_{[2]}\mathcal{C}_{[6]}$,\;$N\mathcal{C}_{[5,3]}$,\; $\mathcal{C}_{[10]}$ & $\hbar^{11}$ & $N\mathcal{C}_{[2]}$, $N^3\mathcal{C}_{[2]}$, $N^5\mathcal{C}_{[2]}$, $N^7\mathcal{C}_{[2]}$, $N^9\mathcal{C}_{[2]}$,\; $N\mathcal{C}_{[2]}^2$, $N^3\mathcal{C}_{[2]}^2$, $N^5\mathcal{C}_{[2]}^2$,\; $N\mathcal{C}_{[4]}$, $N^3\mathcal{C}_{[4]}$, $N^5\mathcal{C}_{[4]}$, $N^7\mathcal{C}_{[4]}$,\; $N\mathcal{C}_{[2]}^3$,\; $N\mathcal{C}_{[2]}\mathcal{C}_{[4]}$, $N^3\mathcal{C}_{[2]}\mathcal{C}_{[4]}$,\; $\mathcal{C}_{[3,3]}$, $N^2\mathcal{C}_{[3,3]}$, $N^4\mathcal{C}_{[3,3]}$,\; $N\mathcal{C}_{[6]}$, $N^3\mathcal{C}_{[6]}$, $N^5\mathcal{C}_{[6]}$,\; $N\mathcal{C}_{[8]}$, $N^3\mathcal{C}_{[8]}$,\;$N\mathcal{C}_{[2]}\mathcal{C}_{[6]}$,\;$N\mathcal{C}_{[4]}^2$,\;
    $\mathcal{C}_{[5,3]}$, $N^2\mathcal{C}_{[5,3]}$,\; $N\mathcal{C}_{[10]}$,\;$\mathcal{C}_{[2]}\mathcal{C}_{[3,3]}$,\;$\mathcal{C}_{[7,3]}$ \\
        \hline
    \end{tabular}
\renewcommand{\arraystretch}{1}
\end{center}
\end{table}
Due to the analyticity of the HOMFLY group factors in $N$ and the conjugation symmetry, odd elements~\eqref{AnalyticZ} are present only in specific combinations. Here are the first ones:
\begin{equation}\label{ZOdd}
{\small
\begin{aligned}
\mathcal{C}^R_{[3,3]}&:=\frac{1}{N}\left(\frac{8\,\mathcal{C}_{[2]}^3+\frac{1}{4}N^4\mathcal{C}_{[3]}^2}{N^2}-\mathcal{C}_{[2]}^2-12\,\mathcal{C}_{[2]} \mathcal{C}_{[4]}\right), \\
\mathcal{C}^R_{[5,3]}&:=\frac{1}{N^3}\left(\frac{\frac{1}{16} \mathcal{C}_{[5]} \mathcal{C}_{[3]} N^6-8\, \mathcal{C}_{[2]}^4}{N^2}+\frac{4}{3} \mathcal{C}_{[2]}^3+36\, \mathcal{C}_{[4]} \mathcal{C}_{[2]}^2\right)-\frac{2\, \mathcal{C}_{[2]}\mathcal{C}_{[3,3]}}{N^2}
-\frac{1}{N}\left(\frac{10}{3} \mathcal{C}_{[2]}^3-\frac{\mathcal{C}_{[2]}^2}{24}-4\, \mathcal{C}_{[4]} \mathcal{C}_{[2]}-30\, \mathcal{C}_{[6]} \mathcal{C}_{[2]}-24\, \mathcal{C}_{[4]}^2\right),
\end{aligned}
}
\end{equation}
The elements~\eqref{AnalyticZeven} and~\eqref{ZOdd} multiplicatively generate the basis elements $\mathcal{C}_{\Delta}^{R}$. In formulas~\eqref{ZOdd} and below we omit the superscript $R$ in Casimir invariants and group factors, where it does not lead to misunderstanding.

One can also make a computer computation of the HOMFLY group factors and notice the \textit{absence of some group factors} that we have not managed to explain in our previous paper~\cite{LST1}, but do explain in this work.

As an example, we provide the HOMFLY group factors up to the 9-th level in Table~\ref{GroupFactors}. The group factors, which are crossed out, are forbidden by some extra symmetry. Notably, these group factors are not allowed to be included even in a special case of $N=0$, which corresponds to the colored Alexander polynomial. After introducing the colored Alexander polynomial, in Subection~\ref{AlScaling}, we demonstrate exactly how the so called 1-hook scaling property restricts the group factors $\mathcal{C}_{[2k]}$ and $\mathcal{C}^k_{[2]}$ to combine.

In Section~\ref{MoreVI}, we use primary group factors for computation of primitive Vassiliev invariants. Thus, we provide primary group factors of the colored HOMFLY polynomial up to the 11-th level in Table~\ref{PrimGroupFactors}.


We have proceeded in calculating the HOMFLY group factors and primary group factors up to the 13-th level. At each level, compare their numbers with the numbers of Vassiliev knot invariants $\dim \mathcal{A}_n$ and primary Vassiliev invariants $\dim \mathcal{P}_n$ (see Table~\ref{dimV&G}). $\dim \mathcal{A}_n$ and $\dim \mathcal{P}_n$ are unknown for $n\geq 13$.

Note that the number of group factors stops to coincide with the number of Vasiliev invariants starting with the 6-th level, and the number of primary group factors -- with the 8-th level. We return to this fact in Subsection~\ref{Vogel}.

\begin{table}[h!]\caption{ Number of Vassiliev invariants and the HOMFLY group factors
} \label{dimV&G}
\begin{center}
\begin{tabular}{|c||c|c|c|c|c|c|c|c|c|c|c|c|c|}
    \hline
    \raisebox{-0.1cm}{$n$} & \raisebox{-0.1cm}{1} & \raisebox{-0.1cm}{2} & \raisebox{-0.1cm}{3} & \raisebox{-0.1cm}{4} & \raisebox{-0.1cm}{5} & \raisebox{-0.1cm}{6} & \raisebox{-0.1cm}{7} & \raisebox{-0.1cm}{8} & \raisebox{-0.1cm}{9} & \raisebox{-0.1cm}{10} & \raisebox{-0.1cm}{11} & \raisebox{-0.1cm}{12} & \raisebox{-0.1cm}{13} \\ [1.5ex]
    \hline
    \hline
    \raisebox{-0.1cm}{$\dim \mathcal{P}_n$} & \raisebox{-0.1cm}{0} & \raisebox{-0.1cm}{1} & \raisebox{-0.1cm}{1} & \raisebox{-0.1cm}{2} & \raisebox{-0.1cm}{3} & \raisebox{-0.1cm}{5} & \raisebox{-0.1cm}{8} & \raisebox{-0.1cm}{12} & \raisebox{-0.1cm}{18} & \raisebox{-0.1cm}{27} & \raisebox{-0.1cm}{39} & \raisebox{-0.1cm}{55} & \raisebox{-0.1cm}{?} \\ [1.5ex]
    \hline
    \raisebox{-0.1cm}{$\dim \varphi_{\mathfrak{sl}_N}^R(\mathcal{D}_n)^{\text{prim}}$} & \raisebox{-0.1cm}{0} & \raisebox{-0.1cm}{1} & \raisebox{-0.1cm}{1} & \raisebox{-0.1cm}{2} & \raisebox{-0.1cm}{3} & \raisebox{-0.1cm}{5} & \raisebox{-0.1cm}{8} & \raisebox{-0.1cm}{\textbf{11}} & \raisebox{-0.1cm}{\textbf{16}} & \raisebox{-0.1cm}{\textbf{22}} & \raisebox{-0.1cm}{\textbf{30}} & \raisebox{-0.1cm}{\textbf{42}} & \raisebox{-0.1cm}{\textbf{53}} \\ [1.5ex]
    \hline
    \hline
    \raisebox{-0.1cm}{$\dim \mathcal{A}_n$} & \raisebox{-0.1cm}{0} & \raisebox{-0.1cm}{1} & \raisebox{-0.1cm}{1} & \raisebox{-0.1cm}{3} & \raisebox{-0.1cm}{4} & \raisebox{-0.1cm}{9} & \raisebox{-0.1cm}{14} & \raisebox{-0.1cm}{27} & \raisebox{-0.1cm}{44} & \raisebox{-0.1cm}{80} & \raisebox{-0.1cm}{132} & \raisebox{-0.1cm}{232} & \raisebox{-0.1cm}{?} \\ [1.5ex]
    \hline
    \raisebox{-0.1cm}{$\dim \varphi_{\mathfrak{sl}_N}^R(\mathcal{D}_n)$} & \raisebox{-0.1cm}{0} & \raisebox{-0.1cm}{1} & \raisebox{-0.1cm}{1} & \raisebox{-0.1cm}{3} & \raisebox{-0.1cm}{4} & \raisebox{-0.1cm}{\textbf{8}} & \raisebox{-0.1cm}{\textbf{11}} & \raisebox{-0.1cm}{\textbf{19}} & \raisebox{-0.1cm}{\textbf{25}} & \raisebox{-0.1cm}{\textbf{39}} & \raisebox{-0.1cm}{\textbf{50}} & \raisebox{-0.1cm}{\textbf{75}} & \raisebox{-0.1cm}{\textbf{95}} \\ [1.5ex]
    \hline
\end{tabular}
\end{center}
\end{table}

\subsection{Combinatorial basis for the HOMFLY group factors }\label{CombConstr}
The algorithm for the group factors \cite{LST1} presented in the previous section allows one to compute a group factor at any level of the perturbative expansion. However, there are some subtleties that are present for odd elements. Here we provide a combinatorial algorithm that computes the HOMFLY group factors in a {\it universal} way, without division into odd and even elements.

Following the ideas discussed in Sections~\ref{colored HOMFLY} and at the beginning of Section~\ref{HOMFLY-GF}, we construct group factors in a two step algorithm.
\begin{enumerate}
    \item We provide a combinatorial description of properly {\it analytically continued basis in the space of $\mathfrak{sl}_N$ Casimir invariants}, which is actually just a $N$-deformed algorithm from~\cite{Mishnyakov2020khb}. Basis elements $\mathcal{G}^R_{\Lambda}$ are enumerated by Young diagrams $\Lambda = [\Lambda_1, \Lambda_1, \Lambda_3, \ldots, \Lambda_{l(\Lambda)}]$ with equal lengths of the first and the second row.
    \item Not all $\mathfrak{sl}_N$ Casimir invariants $\mathcal{G}^R_{\Lambda}$ are included in the perturbative expansion, there are special {\it selection rules}. The rules are described via the symmetries and properties of the colored HOMFLY polynomials.
\end{enumerate}

We start with an explicit combinatorial description of a specific basis $\mathcal{G}_{\Lambda}$ in the space of $\mathfrak{sl}_N$ Casimir eigenvalues. In accordance with $\mathfrak{sl}_N$ representation theory, the number of $\mathfrak{sl}_N$ Casimir invariants of order $n$ is
\begin{equation}
\label{dim}
    \dim \mathcal{G}_{|\Lambda| = n} = p(n) - p(n - 1)\,,
\end{equation}
where $p(n)$ is the number of Young diagrams with $n$ boxes. The basis elements $\mathcal{G}_{\Lambda}$ are enumerated by Young diagrams with equal lengths of the first and the second row, i.e. $\Lambda_1 = \Lambda_2$. The number of such Young diagrams is exactly the dimension \eqref{dim}.
We present the formula for the basis elements:
\begin{equation}
    \label{combcasimir}
    \boxed{
        \mathcal{G}^R_{\Lambda} = \sum_{\substack{|\Delta| = |\Lambda| \\ \Delta \geqslant \Lambda}} \mu^{\Lambda}_{\Delta} \ \xi_{\Delta}^{\Lambda} \ Q^R_{\Delta}\,,
        }
    \end{equation}
where we introduce:
\begin{equation}
     \xi^{\Lambda}_{\Delta} := \frac{(-1)^{\Delta_1 - \Lambda_1} \,  N^{l(\Lambda) - l(\Delta)}} {\prod_i \Delta_i !}\,, \hspace{10mm} Q_{\Delta} :=  \prod_{i = 1}^{l(\Delta)} \left( C_{\Delta_i} + \theta^N_{\Delta_i} \right) - \prod_{i = 1}^{l(\Delta)} \theta^N_{\Delta_i}\,, \hspace{10mm} \theta_k^N = \sum_{i = 1}^N \left( - i + 1/2 \right)^k. 
\end{equation}

The non-negative integer coefficients $\mu^{\Lambda}_{\Delta}$ are determined from the translational invariance of $\mathfrak{sl}_N$ Casimir invariants. Translations act on the Young diagrams by the rule: $R_i \rightarrow R_i + \delta R$. This relation corresponds to the following fact from $\mathfrak{sl}_N$ representation theory: Young diagrams $[R_1 + \delta R, R_2 + \delta R, \ldots, R_N + \delta R]$ and $[R_1, R_2, \ldots, R_N]$ correspond to the same irreducible representation.

We have invented a combinatorial algorithm that provides an answer for the coefficients $\mu^{\Lambda}_{\Delta}$. We construct a weighted directed graph $\Gamma_{\Lambda}$. The vertices are diagrams $\Delta$ from sum $(\ref{combcasimir})$. The graph has levels from left to right according to the number of boxes in the first row, namely, the diagram $\Lambda$ is at the left vertex.

The number of outgoing edges $e$ of each vertex is defined to be the number of \textbf{corner boxes.} The corner is defined as follows: it has adjacent boxes to the top, but does not have adjacent boxes to the right and to the bottom. 

\begin{figure}[h!]
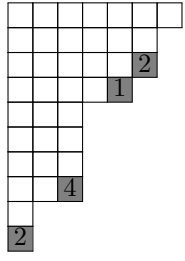

     \begin{center}
    \ytableausetup{boxsize = 0.9em}
        \begin{ytableau}
        \ & \ & \ & \ & \ & \ & \ \\
        \ & \ & \ & \ & \ & \  \\
        \ & \ & \ & \ & \ & *(gray) 2 \\
        \ & \ & \ & \ & *(gray) 1 \\
        \ & \ & \\
        \ & \ & \\
        \ & \ & \\
        \ & \ & *(gray) 4 \\
        \ \\
        *(gray) 2\\ 
        \end{ytableau}
\end{center}
    \caption{Example of valences of corner boxes}
    \label{fig:val}
\end{figure}

We also define a \textbf{valence} of a corner box. It is equal to the number of the diagram's rows that have the same length as the row that contains the corner box. In Fig.~\ref{fig:val}, we provide an example of a diagram to demonstrate the new definitions.
In this figure we highlight the corner boxes and put the valences into them.

As has been mentioned above, the outgoing edges and corner boxes of a vertex-diagram are in the correspondence. An edge connects two diagrams and is weighted by a valence of a corner box $w_e$. A diagram at the head of an edge is obtained from a diagram at the tail as follows: a corner box is cut and glued to the first row of the diagram. In the examples below, note that the rightmost diagram is a one-row diagram.

\textbf{The answer for $\mu_{\Delta}^{\Lambda}$ is given by the sum of weights over paths from $\Lambda$ to $\Delta$ in the graph  $\Gamma_{\Lambda}$:}
\begin{equation}
    \boxed{\mu_{\Delta}^{\Lambda} = \sum_{\substack{\text{paths} \\ \Lambda \rightarrow \Delta}} \ \prod_{e \, \in \, \text{path}} w_e}
\end{equation}
If the graph $\Gamma_{\Lambda}$ does not contain the vertex $\Delta$, there is no suitable path and $\mu_{\Delta}^{\Lambda} = 0$. We provide some examples:
    \begin{equation}\nonumber
\ytableausetup{boxsize = 0.6em}
    \begin{diagram}
        \node{ \begin{ytableau}
                \ \\
                *(gray)$\scriptsize{2}$ \\
               \end{ytableau}} \arrow{e,t}{2}
        \node{ \begin{ytableau}
                \ & \ \\
               \end{ytableau}}
    \end{diagram}
\end{equation}
$$
\mu_{\text{\tiny{[2]}}}^{\text{\tiny{[1,1]}}} = 2
$$
\begin{equation}\label{G[1,1]}
\boxed{
    \mathcal{G}_{[1,1]} = Q_{[1,1]} - N Q_{[2]}
    }
\end{equation}
    \begin{equation}\nonumber
\ytableausetup{boxsize = 0.6em}
    \begin{diagram}
        \node{ \begin{ytableau}
                \ \\
                \ \\
                *(gray)$\scriptsize{3}$ \\
               \end{ytableau}} \arrow{e,t}{3}
        \node{ \begin{ytableau}
                \ & \ \\
                *(gray)$\scriptsize{1}$  \\
               \end{ytableau}} \arrow{e,t}{1}
        \node{ \begin{ytableau}
                \ & \ & \ \\
               \end{ytableau}}
    \end{diagram}
\end{equation}
$$
\mu_{\text{\tiny{[2,1]}}}^{\text{\tiny{[1,1,1]}}} = 3 \hspace{10mm}
\mu_{\text{\tiny{[3]}}}^{\text{\tiny{[1,1,1]}}} = 3 \cdot 1
$$
\begin{equation}\label{G[1,1,1]}
    \boxed{
    \mathcal{G}_{[1,1,1]} = Q_{[1,1,1]} - 3 \frac{N}{2} Q_{[2,1]} + 3 \frac{N^2}{3!} Q_{[3]}
    }
\end{equation}
    \begin{equation}\nonumber
\ytableausetup{boxsize = 0.6em}
    \begin{diagram}
        \node{}
        \node{ \begin{ytableau}
                \ & \ & \ & \ \\
                \ & \ & *(gray) $\scriptsize{1}$ \\
                *(gray) $\scriptsize{1}$ \\
               \end{ytableau}} \arrow{e,t}{1} \arrow{sse,t}{1} 
        \node{ \begin{ytableau}
                \ & \ & \ & \ & \ \\
                \ & \ & *(gray) $\scriptsize{1}$ \\
               \end{ytableau}} \arrow{e,t}{1} 
        \node{ \begin{ytableau}
                \ & \ & \ & \ & \ & \ \\
                \ & *(gray) $\scriptsize{1}$ \\
               \end{ytableau}} \arrow{se,t}{1} \\
        \node{ \begin{ytableau}
                \ & \ & \ \\
                \ & \ & *(gray)$\scriptsize{2}$ \\
                \ & *(gray)$\scriptsize{1}$ \\
               \end{ytableau}}  \arrow{ne,t}{1} \arrow{se,t}{2}
        \node{}
        \node{}
        \node{}
        \node{ \begin{ytableau}
                \ & \ & \ & \ & \ & \ & \ \\
                *(gray) $\scriptsize{1}$ \\
               \end{ytableau}} \arrow{e,t}{1} 
        \node{ \begin{ytableau}
                \ & \ & \ & \ & \ & \ & \ & \ \\
               \end{ytableau}} \\ 
        \node{}
        \node{ \begin{ytableau}
                \ & \ & \ & \ \\
                \ & \ \\
                \ & *(gray) $\scriptsize{2}$ \\
               \end{ytableau}} \arrow{e,t}{2} 
        \node{ \begin{ytableau}
                \ & \ & \ & \ & \ \\
                \ & *(gray) $\scriptsize{1}$ \\
                *(gray) $\scriptsize{1}$ \\
               \end{ytableau}} \arrow{e,t}{1} \arrow{nne,t}{1}
        \node{ \begin{ytableau}
                \ & \ & \ & \ & \ & \ \\
                \ \\
                *(gray) $\scriptsize{2}$ \\
               \end{ytableau}} \arrow{ne,t}{2}
    \end{diagram}
\end{equation}
$$
\mu_{\text{\tiny{[4,2,2]}}}^{\text{\tiny{[3,3,2]}}} = 2 \hspace{10mm}
\mu_{\text{\tiny{[4,3,1]}}}^{\text{\tiny{[3,3,2]}}} = 1 \hspace{10mm}
\mu_{\text{\tiny{[5,3]}}}^{\text{\tiny{[3,3,2]}}} = 1 \cdot 1 = 1\hspace{10mm}
\mu_{\text{\tiny{[5,2,1]}}}^{\text{\tiny{[3,3,2]}}} = (1 \cdot 1)  + (2 \cdot 2) = 5 
$$
$$
\mu_{\text{\tiny{[6,2]}}}^{\text{\tiny{[3,3,2]}}} = (2 \cdot 2 \cdot 1) + (1 \cdot 1 \cdot 1) + (1 \cdot 1 \cdot 1) = 6\hspace{10mm}
\mu_{\text{\tiny{[6,1,1]}}}^{\text{\tiny{[3,3,2]}}} = (2 \cdot 2 \cdot 1) + (1 \cdot 1 \cdot 1) = 5
$$
$$
\mu_{\text{\tiny{[7,2,1,1]}}}^{\text{\tiny{[3,3,2]}}} = (1 \cdot 1 \cdot 1 \cdot 1) + (1 \cdot 1 \cdot 1 \cdot 1) + (1 \cdot 1 \cdot 1 \cdot 2) + (2 \cdot 2 \cdot 1 \cdot 1) + (2 \cdot 2 \cdot 1 \cdot 2) =16
$$
$$
\mu_{\text{\tiny{[8]}}}^{\text{\tiny{[3,3,2]}}} = (1 \cdot 1 \cdot 1  \cdot 1\cdot 1) + (1 \cdot 1 \cdot 1 \cdot 1\cdot 1) + (1 \cdot 1 \cdot 1 \cdot 2\cdot 1) + (2 \cdot 2 \cdot 1 \cdot 1\cdot 1) + (2 \cdot 2 \cdot 1 \cdot 2\cdot 1) =16
$$
\begin{align}\label{Gamma[3,3,2]}
\boxed{
\begin{aligned}
    \mathcal{G}_{[3,3,2]} &= \frac{1}{3! 3! 2!} Q_{[3,3,2]} - 2 \frac{1}{4!2!2!} Q_{[4,2,2]}-\frac{1}{4!3!} Q_{[4,3,1]} + 5 \frac{1}{5! 2!} Q_{[5,2,1]} + \frac{N}{5!3!}Q_{[5,3]} - \\
    &-5 \frac{1}{6!} Q_{[6,1,1]} - 6 \frac{N}{6!2!} Q_{[6,2]} + 16 \frac{N}{7!} Q_{[7,1]} - 16 \frac{N^2}{8!} Q_{[8]}
\end{aligned}
}
\end{align}
Note that the obtained group factors must be linear combinations of the already obtained HOMFLY group factors $\mathcal{C}_\Delta$ formed by the multiplicative basis~\eqref{AnalyticZ},~\eqref{AnalyticZeven}. We provide some examples of this correspondence:
\begin{equation}\label{G-C-corr}
\begin{aligned}
    \mathcal{G}_{[1,1]}&=-2\,\mathcal{C}_{[2]}\,, \\
    \mathcal{G}_{[1,1,1]}&=-\frac{N^2}{2}\mathcal{C}_{3}\,, \\
\end{aligned}
\end{equation}
which one can easily obtain knowing~\eqref{G[1,1]},~\eqref{G[1,1,1]}. \\
A typical group factor of a colored HOMFLY has the following form:
\begin{equation}
    \boxed{
    \mathcal{G}^R_{n,m} = N^k \mathcal{G}_{\Lambda}^R
    }
\end{equation}
However, not all choices of $\Lambda$ and $k$ are allowed in the perturbative expansion. We discuss {\it selection rules} for the group factors, that describe which Casimir elements $\mathcal{G}^R_{\Lambda}$ are present in the perturbative expansion.
\begin{itemize}
    \item {\bf Conjugation symmetry}. The conjugation symmetry is defined at a fixed value of $N$. It comes from the representation theory, where a conjugate representation $\overline{R}$ is defined as a complement of a Young diagram $R$ to the rectangular $R_1 \times N$. The conjugation procedure is clear from the example $N=5$ in Fig.~\ref{fig:ConjR}.
\begin{figure}[h!]
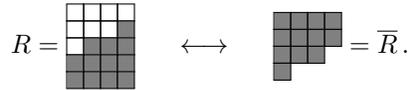

\begin{equation}\nonumber
\ytableausetup{boxsize=0.6em,aligntableaux = center}
R=\begin{ytableau}
 \ & \ & \ & \ \\
 \ & \ & \ & *(gray) \\
 \ & *(gray) & *(gray) & *(gray) \\
  *(gray) & *(gray) & *(gray) & *(gray) \\
  *(gray) & *(gray) & *(gray) & *(gray) \\
\end{ytableau} \hspace{5mm}\longleftrightarrow \hspace{5mm}\begin{ytableau}
 *(gray) & *(gray) & *(gray) & *(gray) \\
 *(gray) & *(gray) & *(gray) & *(gray) \\
 *(gray) & *(gray) & *(gray) \\
 *(gray) \\
\end{ytableau}=\overline{R}\,.
\end{equation}
\caption{Example of a Young diagram and its conjugate}
    \label{fig:ConjR}
\end{figure}\\
The entries of a Young diagram of the conjugate representation $\overline{R}$ are expressed by the following formula:
\begin{equation}
\label{conj transform}
\overline{R}_{i}=R_{1}-R_{N-i+1}\,.
\end{equation}
It is a well-known fact that the colored HOMFLY polynomials for a representation $R$ and its conjugate $\overline{R}$ coincide.
As a direct corollary of formula \eqref{conj transform} and the definition of functions $C_k^{R}$ \eqref{gl Casimirs} one can derive the following transformation rule:
\begin{equation}
\begin{aligned}\label{adjC}
    C_{k}^{\overline{R}}=(-1)^k\sum_{p=0}^{k} \ \epsilon^{p} \ \binom{k}{p}\left(C_{k-p}^R+\theta_{k-p}^N \right)-\theta_k^N,
\end{aligned}
\end{equation}
where remind that the function $\theta_k^{N}=\sum\limits_{i=1}^N\left(-i+\frac{1}{2}\right)^{k}$. Due to the presence of $(-1)^k$ factor the basis elements $\mathcal{C}_{\Lambda}$ obey the following rule:
\begin{equation}
    \boxed{
    |\Lambda|\mod 2 = 0\,.
    }
\end{equation}
\item \textbf{Rank-level duality}. The rank-level duality of the Chern-Simons theory, also called mirror symmetry,  provides the following relation:
\begin{equation}
\mathcal{H}^\mathcal{K}_R(q,a)=\mathcal{H}^\mathcal{K}_{R^T}(q^{-1},a)\,,
\end{equation}
where $R^T$ is the representation obtained by transposing the Young diagram (see Fig.~\ref{fig:R^T} as an example). This property imposes the following condition on group-factors:
\begin{equation}
\label{rank-level-duality}
    \mathcal{G}_{n,m}^{R^T} = (-1)^n \, \mathcal{G}_{n,m}^{R}\Big|_{N\rightarrow-N}\,.
\end{equation}
\begin{figure}[h!]
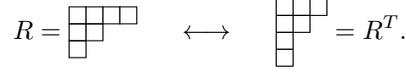

\begin{equation}\nonumber
\ytableausetup{boxsize=0.6em,aligntableaux = center}
R=\ydiagram{4,2,1} \hspace{5mm}\longleftrightarrow \hspace{5mm}\ydiagram{3,2,1,1}=R^T.
\end{equation}
\caption{Example of a Young diagram and its transposed}
    \label{fig:R^T}
\end{figure}
The Casimir eigenvalues of $\mathfrak{gl}_{\infty}$ transform in a simple way under the transposition of the diagram:
\begin{equation}
\label{Ck(RT)}
    C_k^{R^{T}} = (-1)^{k + 1} C_k^{R}\,.
\end{equation}
It means that a basis element $N^k \mathcal{G}_{\Lambda}$ appears in the order $n$ of the perturbative expansion only if the following rule holds:
\begin{equation}
    \boxed{
    n + k + |\Lambda| + l(\Lambda) \mod 2 = 0\,.
    }
\end{equation}
\item \textbf{Genus order}. The genus order restriction comes from the fact that the genus expansion of the colored HOMFLY polynomials is well defined. This property restricts the genus order of the primary group factors. The genus order $g$ is defined for a monomial in the following way:
\begin{equation}
    g\left( N^k C_{\Delta} \right) = k + |\Delta|\,.
\end{equation}
The genus order of a group factor is defined to be the maximal genus order of its components. The genus order of {\it primary} group factor $\mathcal{G}_{n, m}^{R}$ is bounded:
\begin{equation}
     g \left( \mathcal{G}_{n, m}^{R} \right) \leqslant n + 1\,.
\end{equation}
It imposes the following rule for a basis element $N^k \mathcal{G}_{\Lambda}$ at the $n$-th order of the perturbative expansion:
\begin{equation}
    \boxed{
    n \geqslant k + |\Lambda| + l(\Lambda) - 2\,.
    }
\end{equation}
\item {\bf 1-hook scaling property.}
A specialization of the colored HOMFLY polynomial at $a = 1$, the so-called colored Alexander polynomial $\mathcal{A}^{\mathcal{K}}_R(q) := \mathcal{H}_{R}^{\mathcal{K}}(a = 1, q)$ has peculiar property:
\begin{equation}
    \mathcal{A}^{\mathcal{K}}_R (q) = \mathcal{A}_{[1]}^{\mathcal{K}}\left( q^{|R|} \right), \hspace{5mm} \text{where} \hspace{5mm} R = [r,1^{L}].
\end{equation}
It restricts the form of the group factors at even orders of the perturbative expansion. In particular, the following rule holds:
\begin{equation}
\boxed{
    \text{either} \hspace{3mm} \left. \mathcal{G}_{n,m}^{[r,1^L]}\right|_{N=0} \sim (r + L)^n, \hspace{5mm} \text{or} \hspace{5mm} \left. \mathcal{G}_{n,m}^{[r,1^L]}\right|_{N=0} = 0.
    }
\end{equation}
This property gives $n/2 - 1$ additional constraints for every even $n$. For example, Casimir invariant $\mathcal{G}_{[1,1]}$ does not appear at even orders higher than 2 because $\mathcal{G}_{[1,1]}^{[r,1^L]}\Big|_{N=0} \sim (r + L)^{2}$. In contrast, Casimir invariant $\mathcal{G}_{[1,1,1,1]}$ always appears only in combination $12 \mathcal{G}_{[2,2]} + \mathcal{G}_{[1,1,1,1]}$ because it vanishes $\left. \left(12 \mathcal{G}_{[2,2]}^{[r,1^L]} + \mathcal{G}_{[1,1,1,1]}^{[r,1^L]} \right)\right|_{N=0} = 0$.
\end{itemize} 
Note that there is another symmetry of the colored HOMFLY polynomial -- namely, the tug-the-hook symmetry. It does not restrict the HOMFLY group factors, as it manifests in the same way as the translation invariance of $\mathfrak{sl}_N$ representations: $[R_1 + \delta R, R_2 + \delta R, \ldots, R_N + \delta R]=[R_1, R_2, \ldots, R_N]$. In other words, the colored HOMFLY polynomial has the tug-the-hook symmetry by its definition. This fact is proved in an accurate way in~\cite{LST1}.

Imposing all the selection rules, we end up with the following basis of the primary group factors in the low orders (see Table ~\ref{GroupFactorsG}).
\begin{table}[h!]\caption{Primary HOMFLY group factors up to the 6-th order}\label{GroupFactorsG}
\begin{center}
\renewcommand{\arraystretch}{2}
    \begin{tabular}{| m{1em} | m{23em} |} 
        \hline
        $\hbar^2$ & $\mathcal{G}_{[1,1]}$ \\
        \hline 
        $\hbar^3$ & $N\mathcal{G}_{[1,1]}$ \\
        \hline
        $\hbar^4$ & $N^2\mathcal{G}_{[1,1]}$, $\mathcal{G}_{[2,2]}$ \\
        \hline 
        $\hbar^5$ & $N \mathcal{G}_{[1,1]}$, $N^3 \mathcal{G}_{[1,1]}$, $N \mathcal{G}_{[2,2]}$ \\
        \hline
        $\hbar^6$ & $N^2\mathcal{G}_{[1,1]}$, $N^4\mathcal{G}_{[1,1]}$, $12\,\mathcal{G}_{[2,2]}+\mathcal{G}_{[1,1,1,1]}$, $N^2\mathcal{G}_{[2,2]}$, $\mathcal{G}_{[3,3]}$\\
        \hline
    \end{tabular}
\renewcommand{\arraystretch}{1}
\end{center}
\end{table}

\subsection{Group factors associated with Jacobi diagrams}\label{GF-Jacobi}
For future discussion, it is convenient to consider group factors of the colored HOMFLY polynomials in another basis. Namely, we are going to find the group factors $\mathcal{G}_{n,m}$ using the loop expansion~\eqref{LoopExpansionHOMFLY}. We can do this up to the 6-th level, as we know Vassiliev invariants for these levels for many knots~\cite{katlas}. Thus, first, we fix a representation $R$ and for each fixed $n$ solve the system of linear equations on $\mathcal{G}_{n,m}^{R}$:
\begin{equation}
    \sum_{m=1}^{\operatorname{dim}\mathbb{G}_{n}} \mathcal{V}^{\mathcal{K}_i}_{n,m}\mathcal{G}_{n, m}^{R}=\left(\mathcal{H}_{R}^{\mathcal{K}_i}\right)_{n},\quad i=1,\dots,\operatorname{dim}\mathbb{G}_{n},
\end{equation}
where $\left(\mathcal{H}_{R}^{\mathcal{K}_i}\right)_{n}$ is the $\hbar^n$ term in the HOMFLY polynomial $\mathcal{H}_{R}^{\mathcal{K}_i}$.

\begin{figure}[t]
	\center{\includegraphics[scale=0.68]{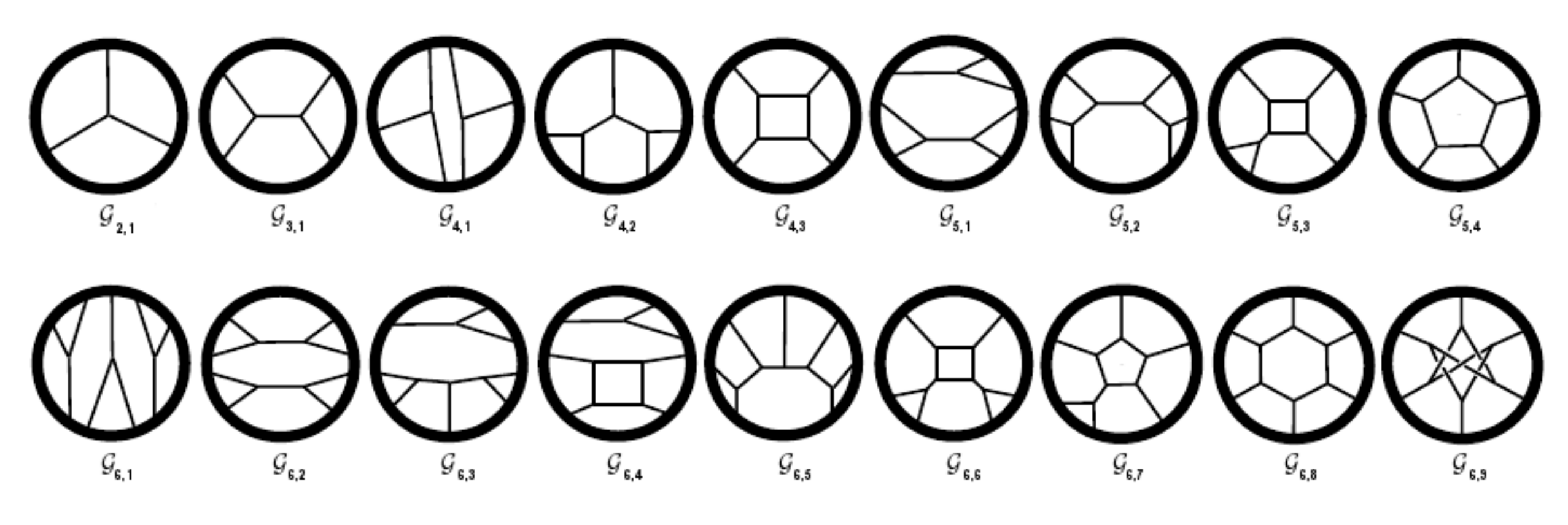}}\caption{Trivalent diagrams}\label{TrivalentDiag}
\end{figure}

We proceed this way for some set of representations $R_\alpha$, and then 
we solve the linear system in unknown $\lambda_{k, l}^{(m)}$:
\begin{equation}
\mathcal{G}_{n, m}^{R_\alpha} =\sum\limits_{|\Delta|\leq n}\mathcal{C}_{\Delta}^{R_\alpha}\sum_{k=0}^{n-|\Delta|} \lambda_{k,l}^{(m)} N^k\in \varphi_{\mathfrak{sl}_N}^{R_\alpha}(\mathcal{D}_n)\subset\mathcal{Z}_n \,,
\end{equation}
for the set of representations $R_\alpha$. Here we utilize the fact that the group factors $\mathcal{G}_{n,m}^{R}$ of the HOMFLY polynomials are linear combinations of the group factors described in Subsection~\ref{Z-GF}, and the knowledge of the HOMFLY polynomials for different knots and representations~\cite{knotebook}. 

\begin{figure}[h!]
	\center{\includegraphics[scale=0.75]{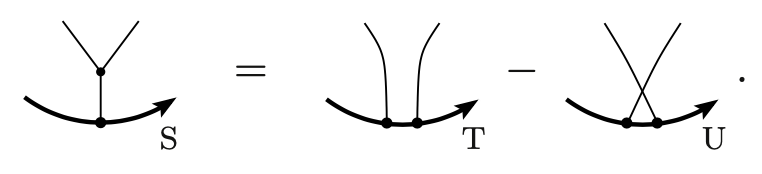}}\caption{STU relation}\label{STU}
\end{figure}

Using this method, we find the group factors $\mathcal{G}_{n,m}$, see Table~\ref{fig:G-Factors}. We know the Vassiliev invariants $\mathcal{V}_{n,m}$ only up to the 6-th order, so we cannot construct group factors in terms of $\mathcal{G}_{n,m}$ furthermore. 
\begin{table}[h!]
\caption{Group factors associated with Jacobi diagrams}\label{fig:G-Factors}
\begin{center}
\renewcommand{\arraystretch}{2}
    \begin{tabular}{| m{1em} | m{21.5em} | m{1em} | m{21.5em} |}
    \hline 
    $\hbar^2$ & $\mathcal{G}_{2,1}=-\frac{1}{2}\mathcal{C}_{[2]}$ & $\hbar^3$ &  $\mathcal{G}_{3,1}=\frac{N}{2}\mathcal{G}_{2,1}=-\frac{N}{4}\mathcal{C}_{[2]}$\\
        \hline 
        $\hbar^4$ & $\mathcal{G}_{4,1}=\left(\mathcal{G}_{2,1}\right)^2=\frac{1}{4}\mathcal{C}_{[2]}^2\,$, $\mathcal{G}_{4,2}=\frac{N^2}{4}\mathcal{G}_{2,1}=-\frac{N^2}{8}\mathcal{C}_{[2]}\,$, $\mathcal{G}_{4,3}=-\frac{N^2}{8}\mathcal{C}_{[2]}+3\mathcal{C}_{[4]}$  
        & $\hbar^5$ & $\mathcal{G}_{5,1}=\mathcal{G}_{2,1}\cdot \mathcal{G}_{3,1}=\frac{N}{8}\mathcal{C}_{[2]}^2\,$, $\mathcal{G}_{5,2}=\frac{N^3}{8}\mathcal{G}_{2,1}=-\frac{N^3}{16}\mathcal{C}_{[2]}\,$, $\mathcal{G}_{5,3}=\frac{N}{2}\mathcal{G}_{4,3}=-\frac{N^3}{16}\mathcal{C}_{[2]}+\frac{3N}{2}\mathcal{C}_{[4]}\,$, $\mathcal{G}_{5,4}=\frac{3N}{4}\mathcal{G}_{4,3}+\frac{4N+N^3}{16}\mathcal{G}_{2,1}=-\frac{N+N^3}{8}\mathcal{C}_{[2]}+\frac{9N}{4}\mathcal{C}_{[4]}$ \\
        \hline 
 \end{tabular}
 \begin{tabular}{| m{1em} | m{46.5em} |}
    \hline 
    $\hbar^6$ &
    $\mathcal{G}_{6,1}=\left(\mathcal{G}_{2,1}\right)^3=-\frac{1}{8}\mathcal{C}_{[2]}^3\,$, $\mathcal{G}_{6,2}=\left(\mathcal{G}_{3,1}\right)^2=\frac{N^2}{16}\mathcal{C}_{[2]}^2\,$, $\mathcal{G}_{6,3}=\mathcal{G}_{2,1}\cdot \mathcal{G}_{4,2}=\frac{N^2}{16}\mathcal{C}_{[2]}^2\,$, $\mathcal{G}_{6,4}=\mathcal{G}_{2,1}\cdot \mathcal{G}_{4,3}=\frac{N^2}{16}\mathcal{C}_{[2]}^2-\frac{3}{2}\mathcal{C}_{[4]} \mathcal{C}_{[2]}\,$, $\mathcal{G}_{6,5}=\frac{N^4}{16}\mathcal{G}_{2,1}=-\frac{N^4}{32}\mathcal{C}_{[2]}\,$, $\mathcal{G}_{6,6}=\frac{N^2}{4}\mathcal{G}_{4,3}=-\frac{N^4}{32}\mathcal{C}_{[2]}+\frac{3N^2}{4}\mathcal{C}_{[4]}\,$, $\mathcal{G}_{6,7}=\frac{N}{2}\mathcal{G}_{5,4}=-\frac{N^2+N^4}{16}\mathcal{C}_{[2]}+\frac{9N^2}{8}\mathcal{C}_{[4]}\,$, $\mathcal{G}_{6,8}=-\frac{N^4+7 N^2}{32} \mathcal{C}_{[2]}+\frac{3}{16}\mathcal{C}_{[2]}^2+\frac{9}{8}\mathcal{C}_{[4]}+\frac{45}{2}\mathcal{C}_{[6]}\,$, $\mathcal{G}_{6,9}=\frac{12 N^2+15}{8}\mathcal{C}_{[4]}-\frac{2 N^4+9 N^2}{32}\mathcal{C}_{[2]}+\frac{5}{16}\mathcal{C}_{[2]}^2-\frac{45}{2}\mathcal{C}_{[6]}$ \\
    \hline
\end{tabular}
\renewcommand{\arraystretch}{1}
\end{center}
\end{table}\\
Due to~\eqref{D-G}, each group factor corresponds to the unframed chord diagram. Yet, in what follows, it is more convenient to work with the so called Jacobi or trivalent diagrams~\cite{chmutov_duzhin_mostovoy_2012}, which vector space is isomorphic to the vector space of unframed chord diagrams due to STU relation (Fig.~\ref{STU}). The $\mathfrak{sl}_N$ weight system associated to a representation $R$ sends Jacobi diagrams straightly into the group factors $\mathcal{G}^R_{n,m}$~\cite{DUNIN_BARKOWSKI_2013}, see Fig.~\ref{TrivalentDiag}.

One can notice that $\mathcal{G}_{6,2}=\mathcal{G}_{6,3}$, that is the manifestation of Vogel's theorem and is discussed in more detail in the next section. Note that this fact has been already remarked at the end of Subsection~\ref{Z-GF} (see Table~\ref{dimV&G} and the comment below).
\\   

\setcounter{equation}{0} 
\section{More Vassiliev invariants}\label{MoreVI}
The possibility of finding Vassiliev invariants was restricted up to the 6-th order by laboriousness
of the calculation of group factors $\mathcal{G}_{n,m}$ by applying the weight system to chord diagrams $\mathcal{D}_{n,m}$~\eqref{D-G}. Our method for decomposition of the HOMFLY polynomials into a linear combination of $\mathfrak{sl}_N$ Casimir invariants and subsequent imposition of the symmetry constraints on them~\cite{LST1}, allows one to get Vassiliev invariants for orders higher than 6th, if we know enough colored HOMFLY polynomials.

Note, however, that in proceeding this way, one can obtain all Vassiliev invariants only up to the 8-th level because starting with the 9-th level primary group factors start to coincide and the weight system of the Lie algebra $\mathfrak{sl}_N$ stops to distinguish the corresponding primary Vassiliev invariants (see Section~\ref{Vogel}). A crucial fact is that now we can calculate Vassiliev invariants at the 11-th level, which can distinguish the mutant knots.

Let us illustrate this new approach and find primary Vassiliev invariants for the knot $5_2$ up to the 8-th level. In order to do this, we use the following expansion:
\begin{equation}\label{PrimVasFind}
    \log(\mathcal{H}_{R}^{\mathcal{K}})=1+\hbar^2\left(v^{\mathcal{K}}_{[2],0}\right)_2\mathcal{C}_{[2]}^{R}+\hbar^3\left(v^{\mathcal{K}}_{[2],1}\right)_3N\mathcal{C}_{[2]}^{R}+\hbar^4\left(\left(v^{\mathcal{K}}_{[2],2}\right)_4N^2\mathcal{C}_{[2]}^{R}+\left(v^{\mathcal{K}}_{[4],0}\right)_4\mathcal{C}_{[4]}^{R}\right)+O(\hbar^5),
\end{equation}
where the primary group factors are taken from Table~\ref{PrimGroupFactors}. Then, we fix a set of representations $R_\alpha$ and at each order $n$ solve a system of linear equations on $\left(v_{\Delta,m}^\mathcal{K}\right)_n$. A set of representations $R_\alpha$, which one uses to obtain primary Vassiliev invariants up to the 11-th level, is provided in Table~\ref{Reps}. We emphasize that the formula \eqref{PrimVasFind} allows one to compute Vassiliev invariants, if a set of the colored HOMFLY polynomials is known. We provide references for explicit formulas for the colored HOMFLY polynomials for a large family of knots and representations:
\begin{itemize}
\item For \textbf{3-strand knots}, there are explicit formulas for $\mathcal{R}$-matrices for representations  $R = [r],  [2,1],[2,2],[3,3]$ \cite{dhara2019multi, mironov2015colored,mironov2016quantum,mironov2016homfly,shakirov2021quantum}, which allow one to compute corresponding HOMFLY polynomials. 
\item For \textbf{arborescent knots}, the same information for the same representations are given in \cite{gu2015note, nawata2013colored,Mironov:2014,mironov2016racah,shakirov2021quantum}. Reference \cite{mironov2017tabulating} can be useful for detailed information on arborescent calculus with explicit examples.
\item Paper \cite{mironov2015towards} explaines how to glue arboresecent blocks into 3-strand braids. It allows one to obtain a really huge family of knots, which contains almost all practical examples.
\item The representation $R=[4,1]$ for \textbf{3-strand knots} were calculated by the same method given in \cite{shakirov2021quantum}. The answer can be found on the cite \cite{knotebook} in  Section \textit{DATA: Racah}.
\end{itemize}
\begin{table}[h!]\caption{A set of representations $R_\alpha$ for~\eqref{PrimVasFind}}\label{Reps}
\begin{center}
\renewcommand{\arraystretch}{2}
    \begin{tabular}{| m{4.5em} | m{12.5em} |} 
        \hline
        $\hbar^2$, $\hbar^3$, $\hbar^4$ & $[1]$ \\
        \hline 
        $\hbar^5$, $\hbar^6$ & $[1]$, $[2]$  \\
        \hline 
        $\hbar^7$ & $[1]$, $[2]$, $[2,1]$ \\
        \hline
        $\hbar^8$ &$[1]$, $[2]$, $[3]$, $[2,1]$ \\
        \hline
        $\hbar^9$ &$[1]$, $[2]$, $[3]$, $[3,1]$ \\
        \hline
        $\hbar^{10}$ &$[1]$, $[2]$, $[3]$, $[2,1]$, $[3,1]$ \\
        \hline
        $\hbar^{11}$ &$[1]$, $[2]$, $[3]$, $[2,1]$, $[3,1]$, $[4,1]$ \\
        \hline
    \end{tabular}
\renewcommand{\arraystretch}{1}
\end{center}
\end{table}
One can also find the colored HOMFLY polynomials in these representations for some knots in~\cite{knotebook}.

Note that in order to shorten the notations we label these non-zero Vassiliev invariants as $v_{n,m}$. Namely, the first Vassiliev invariants correspond to even elements $\mathcal{C}_{[2k]}$~\eqref{AnalyticZeven} and are ordered lexicographically and the rest ones correspond to odd elements $\mathcal{C}_{[2k+1,2l+1]}$~\eqref{ZOdd} and are also labeled in lexicographical order. 
\newpage
For example, for the 7-th level we have: 
\begin{multicols}{4}
\begin{flushleft} $v_{7,1}:=\left(v_{[2],1}^\mathcal{K}\right)_7,$
\end{flushleft} $v_{7,2}:=\left(v_{[2],3}^\mathcal{K}\right)_7,$ \\ \\
$v_{7,3}:=\left(v_{[2],5}^\mathcal{K}\right)_7,$ \\ \\
$v_{7,4}:=\left(v_{[2,2],1}^\mathcal{K}\right)_7,$ \\ \\
$v_{7,5}:=\left(v_{[4],1}^\mathcal{K}\right)_7,$ \\ \\
$v_{7,6}:=\left(v_{[4],3}^\mathcal{K}\right)_7,$ \\ \\
$v_{7,7}:=\left(v_{[6],1}^\mathcal{K}\right)_7,$ \\ \\
$v_{7,8}:=\left(v_{[3,3],0}^\mathcal{K}\right)_7.$
\end{multicols}
List the primary Vassiliev invariants for the knot $\mathcal{K}=5_2$ (Table~\ref{tab:VassInv}).
\begin{table}[h!]\caption{Primary Vassiliev invariants for the knot $\mathcal{K}=5_2$}
    \label{tab:VassInv}
{\small
\begin{center}
    \begin{tabular}{|c||c|c|c|c|c|c|c|c|c|c|c|}
    \hline
    \diagbox[height=1cm,width=1cm]{\raisebox{0.1cm}{$n$}}{\raisebox{-0.07cm}{$m$}} & 1 & 2 & 3 & 4 & 5 & 6 & 7 & 8 & 9 & 10 & 11  \\ 
\hline
\hline 
\raisebox{-0.18cm}{$2$} & \raisebox{-0.18cm}{$-4$} & \ & \ & \ & \ & \ & \ & \ & \ & \ & \ \\ [2ex]
\hline 
\raisebox{-0.18cm}{$3$} & \raisebox{-0.18cm}{$-6$} & \ & \ & \ & \ & \ & \ & \ & \ & \ & \  \\ [2ex]
\hline
\raisebox{-0.18cm}{$4$} & \raisebox{-0.18cm}{$-13$} & \raisebox{-0.18cm}{$44$} & \ & \ & \ & \ & \ & \ & \ & \ & \  \\ [2ex]
\hline
\raisebox{-0.18cm}{$5$} & \raisebox{-0.18cm}{$-7$} & \raisebox{-0.18cm}{$-34$} & \raisebox{-0.18cm}{$222$} & \ & \ & \ & \ & \ & \ & \ & \  \\ [2ex]
\hline
\raisebox{-0.18cm}{$6$} & \raisebox{-0.18cm}{$-70$} & \raisebox{-0.16cm}{$-\frac{1141}{12}$} & \raisebox{-0.18cm}{$186$} & \raisebox{-0.18cm}{$875$} & \raisebox{-0.18cm}{$-1684$} & \ & \ & \ & \ & \ & \  \\ [2ex]
\hline
\raisebox{-0.18cm}{7} & \raisebox{-0.16cm}{$-\frac{739}{12}$} & \raisebox{-0.16cm}{$-\frac{5621}{12}$} & \raisebox{-0.16cm}{$-\frac{5681}{20}$} & \raisebox{-0.18cm}{$16$} & \raisebox{-0.18cm}{$2336$} & \raisebox{-0.18cm}{$3421$} & \raisebox{-0.18cm}{$-17586$} & \raisebox{-0.18cm}{$-262$} & \ & \ & \ \\ [2ex]
\hline
\raisebox{-0.18cm}{8} & \raisebox{-0.16cm}{$-\frac{7135}{6}$} & \raisebox{-0.16cm}{$-\frac{15991}{6}$} & \raisebox{-0.16cm}{$-\frac{321493}{360}$} & \raisebox{-0.18cm}{$-684$} & \raisebox{-0.16cm}{$\frac{5099}{2}$} & \raisebox{-0.18cm}{$19786$} & \raisebox{-0.16cm}{$\frac{159899}{12}$} & \raisebox{-0.18cm}{$-2322$} & \raisebox{-0.18cm}{$-111973$} & \raisebox{-0.18cm}{$135404$} & \raisebox{-0.16cm}{$-\frac{7208}{3}$} \\ [2ex]
\hline
 \end{tabular}   
\end{center}}
\end{table}\\
We emphasize that these Vassiliev invariants are written in the basis of group factors different from~\eqref{LoopExpansionHOMFLY}, so that the $v_{n,m}$ are linear combinations of the $\mathcal{V}_{n,m}$, which one can find in~\cite{katlas}. Vassiliev invariants $v_{n,m}$ for the knot $3_1$ up to the 11-th order and for the knot $5_2$ up to the 10-th order one can find in~\cite{knotebook}. 

\setcounter{equation}{0} 
\section{Distinguishing the Vassiliev invariants}\label{Vogel}
In this section, we discuss an application of the phenomenon of universality from representation theory in physics. Namely, consider a universal parametrization of the adjoint sector of Lie algebras, which was demonstrated in the works of P. Vogel~\cite{VOGEL20111292}. Adjoint representations play a key role in theoretical physics due to the fact that gauge fields belong to them. For example, the gluon fields, which are responsible for the confinement mechanism are described by the adjoint $\mathfrak{su}_{3}$ representation. A possible generalization of the group structures arising in correlators of simpler quantum field theories; for example, the Chern-Simons theory, would allow a better study of correlators in four-dimensional quantum chromodynamics. 

Therefore, we pose a specific problem of generalizing the group structure of the HOMFLY polynomials obtained in~\cite{LST1} by the use of Vogel's parametrization to any simple Lie algebra (Table~\ref{VogelPar} from~\cite{2012}). In this table $\alpha$, $\beta$, $\gamma$ are Vogel's parameters and the parameter $t=\alpha+\beta+\gamma$ is a half of the Casimir element in the adjoint representation for the corresponding Lie algebras. 
\begin{table}[h!]\caption{Vogel’s parameters for simple Lie algebras}\label{VogelPar}
	\center{\includegraphics[scale=0.55]{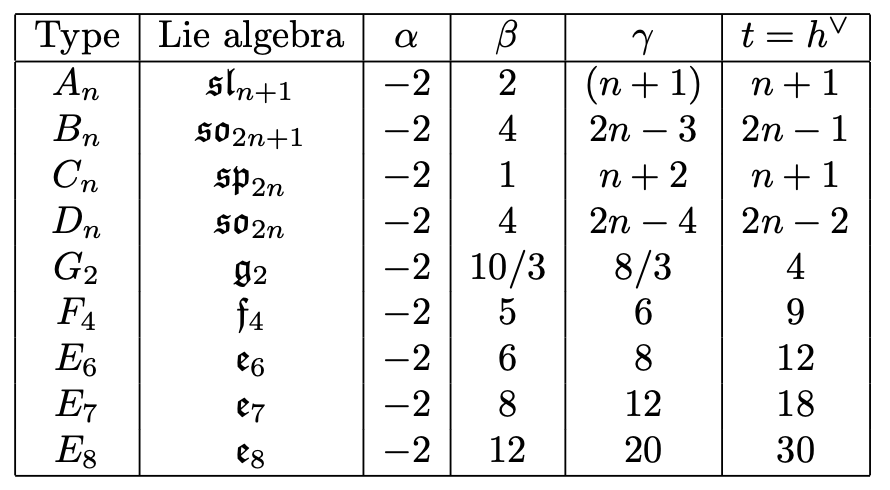}}
\end{table}

Lie algebras' weight systems produce infinite series of examples of Vassiliev invariants. J. Kneissler has shown in~\cite{Kn0} that all invariants up to order 12 come from Lie algebras. However, in general, this is not the case. P. Vogel~\cite{VOGEL20111292} has used the family of Lie superalgebras $D(1, 2, \alpha)$ depending on the parameter $\alpha$; he showed that these algebras produce invariants which cannot be expressed as combinations of invariants coming from Lie algebras. J. Lieberum~\cite{doi:10.1142/S0218216599000420} gave an example of an order 17 closed diagram detected by $D(1, 2, \alpha)$, but not by semisimple Lie algebras' weight systems, and proved that for order $d=15$, there exists a non-zero linear combination of chord diagrams which is turned to zero by all weight systems coming from semisimple Lie algebras. Moreover, there exist Vassiliev invariants that do not come from Lie (super)algebras~\cite{VOGEL20111292, doi:10.1142/S0218216599000420}.
The main technical tool for proving these results is the algebra $\Lambda$ constructed by Vogel.

\begin{figure}[h!]
\begin{minipage}[h]{0.49\linewidth}
\center{\includegraphics[width=0.7\linewidth]{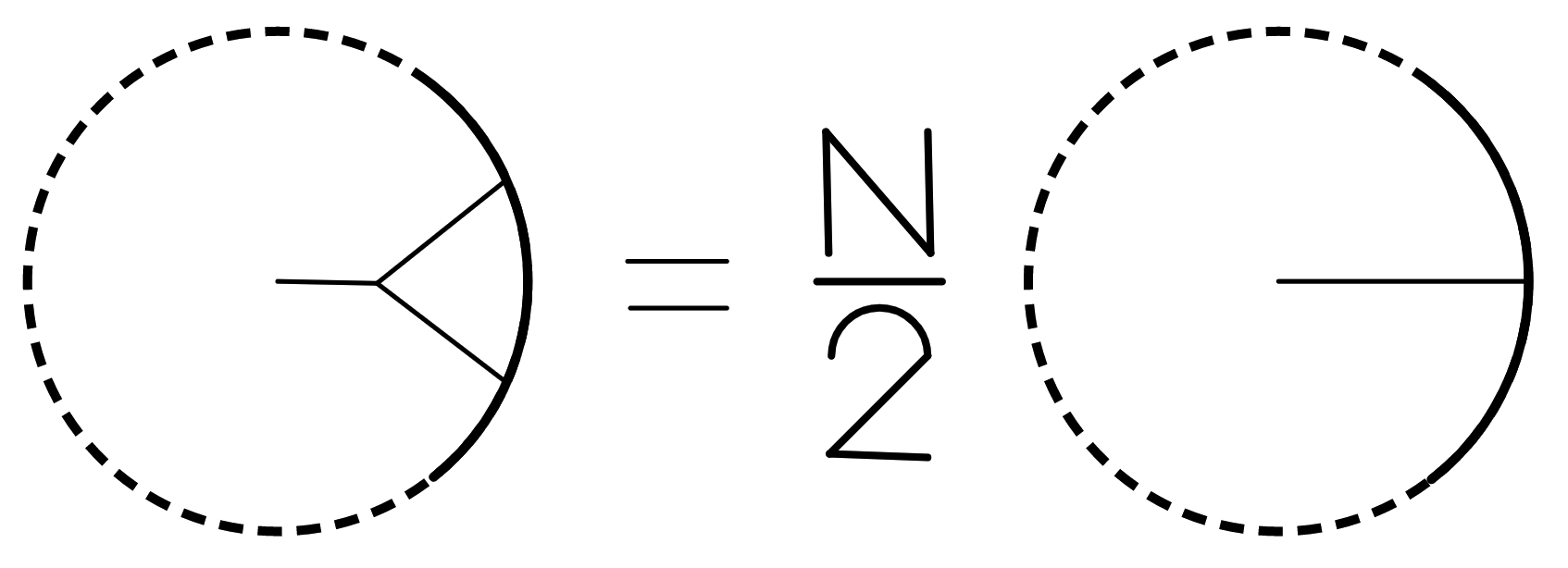}}
\caption{Extra relation}\label{ExtraRel}
\end{minipage}
\hfill
\begin{minipage}[h]{0.49\linewidth}
\center{\includegraphics[width=0.8\linewidth]{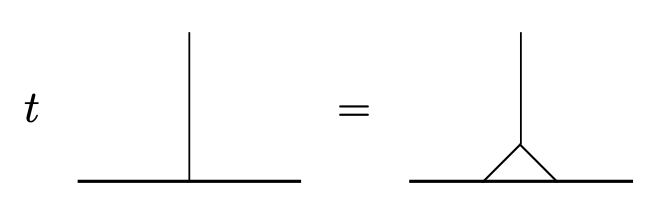}}
\caption{{\small General relation between the Jacobi diagrams}}\label{VogelRel}
\end{minipage}
\end{figure}

In this paper, we find a simple example for the particular $\mathfrak{sl}_N$ weight system at the 6-th level: $\varphi^{R}_{\mathfrak{sl}_N} \left( \mathcal{D}_{6,2} \right)=\varphi^{R}_{\mathfrak{sl}_N} \left( \mathcal{D}_{6,3} \right)$. Namely, knowing the explicit expressions for $\mathcal{G}_{n,m}$ up to the 6-th order (see Subsection~\ref{GF-Jacobi}), we notice that
\begin{equation}\label{GEq}
    \mathcal{G}_{6,2}=\mathcal{G}_{6,3}\,.
\end{equation}
This fact follows from an interesting relation between Jacobi diagrams associated with the $\mathfrak{sl}_N$ weight system (see Fig.~\ref{ExtraRel}), which can be easily obtained from the definition of the Lie algebra weight system and from the $\mathfrak{sl}_N$ commutation relations:
\begin{equation}\label{ChordRelExpl}
    \tr_R\left(\sum\limits_{b,c}f_{abc}T_b T_c\dots\right)=\frac{1}{2} \tr_R\left(\sum\limits_{b,c}f_{abc}[T_b,T_c]\dots\right)=\frac{1}{2}\tr_R\left(f_{abc}f_{a'bc}T_{a'} \dots \right)=\frac{N}{2}\tr_R(T_a\dots)\,.
\end{equation}
Moreover, due to this property, some of the group factors from the higher levels are connected with the group factors from the lower levels by multiplication by $N/2$: for example, $\mathcal{G}_{3,1}=\frac{N}{2}\mathcal{G}_{2,1}$. 

Actually, equality~\eqref{GEq} is generalized to all semisimple Lie algebras by introducing the Vogel parameters $\alpha,\beta,\gamma$. This fact is explained by relation~\eqref{ChordRelExpl} being easily generalized, as $f_{abc}f_{a'bc}=\frac{1}{2}\delta_{aa'}C_2^{\text{adj}}$, and in the Vogel's work~\cite{VOGEL20111292} is drawn as the generalization of Fig.~\ref{ExtraRel} (see Fig.~\ref{VogelRel}). This, however, does not mean that in knot polynomials we are not able to distinguish the Vassiliev invariants $\mathcal{V}_{6,2}$ and $\mathcal{V}_{6,3}$, as they are not primary. Alas, actually, we face the problem of not distinguishing Vassiliev invariants by the $\mathfrak{sl}_N$ weight system on the 8-th level (see the last table in Subsection~\ref{Z-GF}). In fact, it is a much more interesting question if Vogel's parametrization solves this problem, and will be a subject of our future research.

\setcounter{equation}{0} 
\section{$q$-holonomicity of colored Jones polynomials}\label{Jones}
A function of several variables is called holonomic if, roughly speaking, it is determined from finitely many of its values via finitely many linear recursion relations with polynomial coefficients. Zeilberger was the first to notice that the abstract notion of holonomicity can be applied to verify, in a systematic and computerized way, combinatorial identities among special functions. The colored Jones function of a knot is a sequence of Laurent polynomials. It was shown by Stavros Garoufalidis and Thang T Q Le~\cite{Garoufalidis_2005} that such sequences are $q$-holonomic, that is, they satisfy linear $q$-difference equations with coefficients Laurent polynomials in $q$ and $q^n$, where $n$ is the dimension of an irreducible representation of $\mathfrak{sl}_2$.

In this section we discuss the phenomenon of $q$-holonomicity of the colored Jones polynomials. Using a concrete example of the torus knots $T[2,2k+1]$, we consider a method of constructing recursive relations for the Jones polynomials:
\begin{equation}
    J^\mathcal{K}_{r}=\mathcal{H}_{[r]}^\mathcal{K}\left(q, q^2\right).
\end{equation}
Note that the Jones polynomials are the HOMFLY polynomials for $\mathfrak{sl}_2$, so they depend only on one group variable $r:=R_1$. Therefore, the perturbative expansion~\eqref{LoopExpansionHOMFLY} for the Jones polynomial takes the following form:
\begin{equation}\label{JExp}
    J^\mathcal{K}_{r}=1+\hbar^2 \mathcal{V}^{\mathcal{K}}_{2,1}\mathcal{G}_{2,1}^{r}+\hbar^3 \mathcal{V}^{\mathcal{K}}_{3,1}\mathcal{G}_{3,1}^{r}+\hbar^4\left(\mathcal{V}^{\mathcal{K}}_{4,1}\mathcal{G}_{4,1}^{r}+\mathcal{V}^{\mathcal{K}}_{4,2}\mathcal{G}_{4,2}^{r}+\mathcal{V}^{\mathcal{K}}_{4,3}\mathcal{G}_{4,3}^{r}\right)+\dots,
\end{equation}
where the group factors $\mathcal{G}^r_{n,m}$ are polynomials in $r$:
\begin{equation}
\begin{aligned}
    \mathcal{G}_{2,1}^{r}&=\mathcal{G}_{3,1}^{r}=\mathcal{G}_{4,2}^{r}=-\frac{1}{4}r(r+2)\,, \\
    \mathcal{G}_{4,3}^{r}&=2\,\mathcal{G}_{4,1}^{r}=\frac{1}{8}r^2(r+2)^2\,,
\end{aligned}
\end{equation}
etc.

As has been discussed above, we would like to derive a recursive relation for an arbitrary knot $\mathcal{K}$ in the form
\begin{equation}
    a_0(q,r,\mathcal{K})+\sum_{i=1}^k a_i(q,r,\mathcal{K})J^\mathcal{K}_{r+i-1}=0\,,
\end{equation}
where $a_i$ are Laurent polynomials in $q$ and $q^r$ with finite amount of terms.

Let us explicitly describe a method of constructing recursive relations for the Jones polynomials of the torus knot $T[2,2k+1]$. Motivated by the work~\cite{Garoufalidis_2005}, where the three-term relation for the trefoil was derived:
\begin{equation}\label{trefoil}
    J_{\mathcal{K}}(n)=\frac{q^{n-1}+q^{4-4 n}-q^{-n}-q^{1-2 n}}{q^{1 / 2}\left(q^{n-1}-q^{2-n}\right)} J_{\mathcal{K}}(n-1)+\frac{q^{4-4 n}-q^{3-2 n}}{q^{2-n}-q^{n-1}} J_{\mathcal{K}}(n-2)\,,
\end{equation}
for arbitrary $k$ we find the recursion formula in the same form:
\begin{equation}\label{3-term}
    A(q,r,k)\mathcal{J}_{r+1}+B(q,r,k)\mathcal{J}_{r}+C(q,r,k)\mathcal{J}_{r-1}=0\,,
\end{equation}
where we introduce the normalized Jones polynomial $\mathcal{J}_r:=J^\mathcal{K}_{r}\,\text{qdim}(r)$ to simplify calculations; $\text{qdim}(r)=\frac{q^{r+1}-q^{-r-1}}{q-q^{-1}}$ is the quantum dimension of the representation $[r]$. 

Substitute in~\eqref{3-term} the coefficients $A(q,r,k)$ and $C(q,r,k)$ in the form conformed with~\eqref{trefoil}:
\begin{equation}
 \begin{aligned}
    A(q,r,k)&=q^{a_1(k)r+a_2(k)}-q^{a_3(k)r+a_4(k)}, \\
    C(q,r,k)&=q^{c_1(k)r+c_2(k)}-q^{c_3(k)r+c_4(k)},
\end{aligned}   
\end{equation}
and the Jones polynomials as an $\hbar$-expansion~\eqref{JExp}. Get the coefficient $B(q,r,k)$ in the form of a series in $\hbar$ and $r$:
\begin{equation}\label{BSeries}
    B(q,r,k)=\sum\limits_{i,j}b_{ij}\hbar^i r^j.
\end{equation}
We would like this series to be the following polynomial in $q$ (which is in correspondence with~\eqref{trefoil}):
\begin{equation}\label{BPolynomial}
    B(q,r,k)=q^{b_1(k)r+b_2(k)}+q^{b_3(k)r+b_4(k)}-q^{b_5(k)r+b_6(k)}-q^{b_7(k)r+b_8(k)}.
\end{equation}
Then, one can write down the matching conditions between~\eqref{BSeries} and~\eqref{BPolynomial} explicitly for several torus knots $T[2,2k+1]$ and find the unknown coefficients $a_i$, $b_i$, $c_i$. Noticing the linearity of these coefficients in $k$, we find a general formula:
\begin{equation}\label{T[2,2k+1]}
\begin{aligned}
    &q(q^{2r+1}-q^{-2r-1})\mathcal{J}_{r+1}+(q^{-2r(2k+1)-6k-2}(q^{2r+1}-q^{-2r-1})-q^{-2k}(q^{2r+3}-q^{-2r-3}))\mathcal{J}_r+\\
    &+(q^{-4r(k+1)-4(k+1)}-q^{-4kr-4k+2})\mathcal{J}_{r-1}=0\,,
\end{aligned}
\end{equation}
which, however, was already known, but in a slightly different form~\cite{Hikami_2004}.

Despite the fact that the recursive relation for the Jones polynomials of the torus knot $T[2,2k+1]$~\eqref{T[2,2k+1]} is not new, it provides an example of a general method for finding recursive relations for the colored Jones polynomials. Moreover, starting from the recursive formula~\eqref{T[2,2k+1]}, one can obtain relations between Vassiliev invariants for the torus knots $T[2,2k+1]\,$:
\begin{equation}\label{VasInvT[2,2k+1]}
\begin{aligned}
    \mathcal{V}^{\mathcal{K}}_{2,1}&=2 k (k+1)\,; \\
    \mathcal{V}^{\mathcal{K}}_{3,1}&= \frac{4}{3} k (k+1) (2 k+1)\,; \\
    \mathcal{V}^{\mathcal{K}}_{4,2}&=\frac{1}{3} k (k+1) \left(14 k^2+14 k+3\right),\\ \mathcal{V}^{\mathcal{K}}_{4,3}&=\frac{1}{3} k (k+1) \left(2 k^2+2 k+1\right); \\
    \mathcal{V}^{\mathcal{K}}_{5,2}&=\frac{2}{27} \left(9 \mathcal{V}^{\mathcal{K}}_{5,3}+108 k^5+270 k^4+232 k^3+78 k^2+8 k\right),\\ \mathcal{V}^{\mathcal{K}}_{5,4}&=\frac{2}{135} \left(-45 \mathcal{V}^{\mathcal{K}}_{5,3}+144 k^5+360 k^4+340 k^3+150 k^2+26 k\right); \\
    \mathcal{V}^{\mathcal{K}}_{6,5}&=\frac{1}{540} \left(360 \mathcal{V}^{\mathcal{K}}_{6,6}-720 \mathcal{V}^{\mathcal{K}}_{6,8}+9488 k^6+28464 k^5+31980 k^4+16520 k^3+3815 k^2+299 k\right), \\ 
    \mathcal{V}^{\mathcal{K}}_{6,7}&=\frac{1}{135} \left(-90 \mathcal{V}^{\mathcal{K}}_{6,6}-360 \mathcal{V}^{\mathcal{K}}_{6,8}+976 k^6+2928 k^5+3420 k^4+1960 k^3+541 k^2+49 k\right), \\
    \mathcal{V}^{\mathcal{K}}_{6,9}&=\frac{1}{45} \left(45 \mathcal{V}^{\mathcal{K}}_{6,8}+32 k^6+96 k^5+120 k^4+80 k^3+30 k^2+6 k\right).
\end{aligned}
\end{equation}
Here we impose conditions on the Vassiliev invariants~\cite{Torus}:
\begin{equation}\label{VassCond}
\begin{array}{ll}
\begin{aligned}
\mathcal{V}^{\mathcal{K}}_{4,1}&=\frac{1}{2} (\mathcal{V}^{\mathcal{K}}_{2,1})^{2}\,, & \mathcal{V}^{\mathcal{K}}_{6,2}=\frac{1}{2} (\mathcal{V}^{\mathcal{K}}_{3,1})^{2}\,, \\
\mathcal{V}^{\mathcal{K}}_{5,1}&=\mathcal{V}^{\mathcal{K}}_{2,1} \mathcal{V}^{\mathcal{K}}_{3,1}\,, & \mathcal{V}^{\mathcal{K}}_{6,3}=\mathcal{V}^{\mathcal{K}}_{2,1} \mathcal{V}^{\mathcal{K}}_{4,2}\,, \\
\mathcal{V}^{\mathcal{K}}_{6,1}&=\frac{1}{6} (\mathcal{V}^{\mathcal{K}}_{2,1})^{3}\,, & \mathcal{V}^{\mathcal{K}}_{6,4}=\mathcal{V}^{\mathcal{K}}_{2,1} \mathcal{V}^{\mathcal{K}}_{4,3}\,,
\end{aligned}
\end{array}
\end{equation}
so that in~\eqref{VasInvT[2,2k+1]} only primitive Vassiliev invariants are listed.

An interesting feature is that from the recursive relation~\eqref{T[2,2k+1]} one can obtain the explicit expressions for Vassiliev invariants for an arbitrary torus knot $T[2,2k+1]$ up to the level 4~\eqref{VasInvT[2,2k+1]}. Note that the obtained formulas coincide with earlier studies~\cite{Sleptsov_2016,Torus}.

The other side of this problem is that one, on the contrary, can fix the Vassiliev invariants and from the recursive formula~\eqref{T[2,2k+1]} find the relations on the Casimir eigenvalues of $\mathfrak{sl}_N$. To illustrate this phenomenon, let us write down the first two difference relations. The coefficient at $\hbar^3$ gives:
\begin{equation}\label{CasRel3}
k (k+1) \left[-2r(2r+3)\mathcal{G}_{2,1}^{r-1}+2(2r^2+5r+2)\mathcal{G}_{2,1}^{r+1}-4(r+1)\mathcal{G}_{2,1}^{r}+4 r^3+12 r^2+11 r+3\right]=0\,.
\end{equation}
The coefficient at $\hbar^4$ after imposing~\eqref{CasRel3} gives:
\begin{equation}
\begin{aligned}
    &-\frac{1}{3} k (k+1) (2 k+1) \big[-r(18+57r+48r^2+12r^3)(\mathcal{G}_{2,1}^{r-1}-\mathcal{G}_{2,1}^{r+1})-\\
    &-4r(2r+3)\mathcal{G}_{3,1}^{r-1}+4(2r^2+5r+2)\mathcal{G}_{3,1}^{r+1}-8 (r+1)\mathcal{G}_{3,1}^{r}+\\
    &+12 r^5+60 r^4+113 r^3+99 r^2+40 r+6\big]=0\,.
\end{aligned}
\end{equation}
A notable fact is that difference relations on the $\mathcal{G}_{n,m}$ do not depend on the parameter $k$.

\setcounter{equation}{0} 
\section{Consequences for Alexander polynomials} \label{Alexander}
The studied HOMFLY group structure~\eqref{GroupFactors} gives series of implications for a special value of the colored HOMFLY polynomial -- the so called Alexander polynomial:
\begin{equation}\label{AlexanderDef}
    \mathcal{A}_{R}^{\mathcal{K}}(q)=\mathcal{H}_{R}^\mathcal{K}\left(q, q^0\right).
\end{equation}
In Subsection~\ref{NewSymmAlexander}, we discuss the existence of a new symmetry of Alexander polynomials which follows from the symmetry constraints imposed by the conjugation symmetry of the HOMFLY polynomials. Then, in Subsection~\ref{AlToHOMFLY}, we introduce ways of inducing the parameter $a=q^N$ into the Alexander polynomial in order to deform it into the HOMFLY polynomial. And, finally, in Subsection~\ref{AlScaling} we describe a generalization of the previously known 1-hook scaling property to an arbitrary representation $R$.

\subsection{A novel symmetry of Alexander polynomials}\label{NewSymmAlexander}
The group factors of Alexander polynomials were studied in~\cite{Mishnyakov2020khb}. Actually, polynomials $Y_\Lambda$ are not presented on the odd levels, or in other words, $\mathbf{Y}_{2k+1}=\varnothing$, as it must be from our study of the group factors of the colored HOMFLY polynomials. Currently, known symmetries of Alexander polynomials (tug-the-hook symmetry, rank-level duality, and 1-hook scaling property) do not explain this vanishing of $Y_\Lambda$ on the odd levels, and the conjugation symmetry is not valid for $N=0$. Therefore, Alexander polynomials must have some still undiscovered symmetries, which will be studied in further work.

To illustrate the described effect and establish the correspondence between group factors of Alexander polynomials in the basis of $Y_\Lambda$ and $\mathcal{C}_\Delta$, we write down first 8 orders in Table~\ref{AlexanderGF}.
\begin{table}[h!]\caption{Alexander group factors}\label{AlexanderGF}
{\scriptsize 
\begin{center}
\begin{doublespace}
    \begin{tabular}{| c | c | c | c | c | c | c | c |}
         \hline
         $\mathbf{Y}_1$ & $\mathbf{Y}_2$ & $\mathbf{Y}_3$ & $\mathbf{Y}_4$ & $\mathbf{Y}_5$ & $\mathbf{Y}_6$ & $\mathbf{Y}_7$ & $\mathbf{Y}_8$ \\
         \hline
         $\cancel{Y_{[1]}}$ & $Y_{[1,1]}=\mathcal{C}_{[2]}\Big|_{N=0}$ & $\cancel{Y_{[1]}^3}$ & $Y_{[1,1]}^2$ & $\cancel{Y_{[1]}^5}$ & $Y_{[1,1]}^3$ & $\cancel{Y_{[1]}^7}$ & $Y_{[1,1]}^4$ \\
         & & & $Y_{[2,2]}=\mathcal{C}_{[4]}\Big|_{N=0}$ & $\cancel{Y_{[2,2]}Y_{[1]}}$ & $Y_{[2,2]}Y_{[1,1]}$ & $\cancel{Y_{[2,2]}Y_{[1]}^3}$ & $Y_{[2,2]}Y_{[1,1]}^2$ \\
         & & & & & $Y_{[2,2,2]}=\mathcal{C}_{[3,3]}\Big|_{N=0}$ & $\cancel{Y_{[2,2,2]}Y_{[1]}}$ & $Y_{[2,2,2]}Y_{[1,1]}$ \\
         & & & & & $Y_{[3,3]}=\mathcal{C}_{[6]}\Big|_{N=0}$ & $\cancel{Y_{[3,3]}Y_{[1]}}$ & $Y_{[3,3]}Y_{[1]}^2$ \\
         & & & & & & & $Y_{[2,2]}^2$ \\
         & & & & & & & $Y_{[3,3,2]}=\mathcal{C}_{[5,3]}\Big|_{N=0}+\frac{1}{12}\mathcal{C}_{[3,3]}\Big|_{N=0}$ \\
         & & & & & & & $Y_{[4,4]}=\mathcal{C}_{[8]}\Big|_{N=0}$ \\ [1ex]
         \hline
    \end{tabular}
\end{doublespace}
\end{center}
}
\end{table}

\subsection{Deformation from Alexander to HOMFLY}\label{AlToHOMFLY}
The HOMFLY polynomials generalizes the Alexander polynomials~\eqref{AlexanderDef}. In other words
\begin{equation}
    \mathcal{H}_{R}^\mathcal{K}\left(q, q^N\right)\overset{ N=0}{\xrightarrow{\hspace*{1cm}}}\mathcal{A}_{R}^\mathcal{K}\left(q\right).
\end{equation}
It is tempting to search for reverse rules, i.e. to find out the so called $N$-deformation from the Alexander polynomials to the HOMFLY polynomials:
\begin{equation}
    \mathcal{A}_{R}^\mathcal{K}\left(q\right)\underset{\text{rules}}{\overset{N}{\xrightarrow{\hspace*{1cm}}}}\mathcal{H}_{R}^\mathcal{K}\left(q, q^N\right)\,.
\end{equation}
Here we present some details for analytical $N$-deformation of the Alexander group structure to the HOMFLY group structure. We have managed to find explicitly an analytical $N$-deformation of the Alexander group structure only for primary group factors from even levels:
\begin{equation}
    \mathcal{Y}_{[2n]}\Big|_{C_k\rightarrow\, C_k+\,\theta_k}-\mathcal{Y}_{[2n]}\Big|_{C_k\rightarrow\,\theta_k}+\frac{N C_{2k}}{(2k)!}=\mathcal{C}_{[2n]}\,,
\end{equation}
where $\mathcal{Y}_{\Lambda}$ are the basis group factors for Alexander polynomials, namely
\begin{equation}
    \mathcal{Y}_{[2n]}=\mathcal{C}_{[2n]}\Big|_{N=0}\,.
\end{equation}
In Subsection~\ref{CombConstr}, we have also deformed the combinatorial algorithm obtained in~\cite{Mishnyakov2020khb} for constructing the Alexander group factors to the combinatorial algorithm for an arbitrary $N$.

\subsection{Alexander scaling relation}\label{AlScaling}
The colored Alexander polynomial displays a property with respect to $R$~\cite{Itoyama_2012,Mironov_2018}:
\begin{equation}\label{1-hook-SP}
\mathcal{A}_{R}^{\mathcal{K}}(q)-\mathcal{A}_{[1]}^{\mathcal{K}}\left(q^{|R|}\right)=0, \quad \text { where } R=\left[r, 1^{L}\right],
\end{equation}
which holds only for the representations corresponding to 1-hook Young diagrams. 

This 1-hook scaling property has an interesting feature. It restricts the group factors of~\eqref{1-hook-SP} to be equal to the Hirota equations on the Kadomtsev-Petviashvili tau-function~\cite{Mishnyakov:2019otq, MIRONOV2018268}. But note that this correspondence has been established only for symmmetric representations.

We would like to generalize this relation to an arbitrary representation. This general scaling relation is also conjecturally connected with some tau-function or a class of tau-functions. The possibility of uncovering such a correspondence with integrable or non-integrable hierarchies is an interesting question for future researches. 

Note that $\mathcal{C}_\Delta$ in the case of Alexander polynomials (for $N=0$ all $\theta_k^{N}\equiv 0$ in~\eqref{AnalyticZ}) become homogeneous polynomials in $C_k$ of order $|\Delta|$, i.e. $\mathcal{C}_\Delta\rightarrow \lambda^{|\Delta|}\mathcal{C}_\Delta$ if $C_k\rightarrow \lambda^k C_k$. 
Due to this fact we discuss the right way of scaling $C_k\,$. From~\eqref{TTHC} for the case of the Alexander polynomial it follows that $C_k\rightarrow\lambda^k C_k$ if
\begin{equation}
    \alpha_i-\frac{1}{2}\rightarrow\lambda\left(\alpha_i-\frac{1}{2}\right)\text{ and }\beta_i-\frac{1}{2}\rightarrow\lambda\left(\beta_i-\frac{1}{2}\right),
\end{equation}
so for convenience we denote a representation in Frobenius notation as $(\vec{\alpha}|\vec{\beta})$ and its scaling as $(\lambda\vec{\alpha}|\lambda\vec{\beta})$.

The Alexander polynomial is expressed in terms of $\mathcal{C}_\Delta$ in the following way: 
\begin{equation}\label{AlExp}
\begin{aligned}
   \mathcal{A}_{R}^{\mathcal{K}}(\hbar)&=1+\hbar^2 v_{2,1}\mathcal{C}_{[2]}+\hbar^4 \left(\frac{1}{2}v^2_{2,1}\mathcal{C}_{[2]}^2+v_{4,2}\mathcal{C}_{[4]}\right)+\\
   &+\hbar^6\left(\frac{1}{6}v^3_{2,1}\mathcal{C}_{[2]}^3+v_{2,1}v_{4,2}\mathcal{C}_{[2]}\mathcal{C}_{[4]}+v_{6,5}\,\mathcal{C}_{[6]}\right)+\boxed{\hbar^6v_{6,3}\tilde{\mathcal{C}}_{[4]}}+\boxed{\hbar^7 v_{7,8}\mathcal{C}_{[3,3]}}+O(\hbar^{8}),
\end{aligned}
\end{equation}
where we denote $\tilde{\mathcal{C}}_{[4]}:=\mathcal{C}_{[4]}+\frac{1}{6}\mathcal{C}_{[2]}^2$. We have defined the Vassiliev invariants $v_{n,m}$ in Subsection~\ref{MoreVI}.

One can see that the transformation of the Alexander polynomial
\begin{equation}
   \mathcal{A}^\mathcal{K}_{(\vec{\alpha}|\vec{\beta})}(\hbar)\rightarrow \mathcal{A}^\mathcal{K}_{(\lambda\vec{\alpha}|\lambda\vec{\beta})}\left(\frac{\hbar}{\lambda}\right)
\end{equation}
almost leaves it invariant. It means that, for example, boxed terms in~\eqref{AlExp} change after such a transformation. The remaining unhomogeneous terms are:
\begin{equation}\label{A1}
\begin{aligned}
    \mathbf{A}_1(\lambda_0=1,\lambda)&:=\mathcal{A}^\mathcal{K}_{(\vec{\alpha}|\vec{\beta})}(\hbar)-\mathcal{A}^\mathcal{K}_{(\lambda\vec{\alpha}|\lambda\vec{\beta})}\left(\frac{\hbar}{\lambda}\right)=\\
    &=\hbar^6 v_{6,3}\left(1-\frac{1}{\lambda^2}\right)\tilde{\mathcal{C}}^{(\vec{\alpha}|\vec{\beta})}_{[4]}+ \hbar^7 v_{7,8}\left(1-\frac{1}{\lambda}\right)\mathcal{C}^{(\vec{\alpha}|\vec{\beta})}_{[3,3]}+O(\hbar^8)=O(\hbar^6),
\end{aligned}
\end{equation}
where in round bracket we indicate the scaling parameters.

Note that for symmetric representations~\eqref{A1} is equivalent to 1-hook scaling property~\eqref{1-hook-SP}, and $\mathbf{A}_1$ must vanish. In other words, the linear combinations of $\mathcal{C}$-polynomials, taken in symmetric representations, must be zero. Even elements $\mathcal{C}^{(\vec{\alpha}|\vec{\beta})}_{[2k]}$ do not vanish in 1-hook representations, so actually they must combine with $\left(\mathcal{C}^{(\vec{\alpha}|\vec{\beta})}_{[2]}\right)^{k}$. This exactly corresponds to the phenomenon which we have seen in Table~\ref{GroupFactors}.

For an arbitrary representation, actually, it is also possible to vanish all the remaining terms. Let us redenote $\lambda_1:=\lambda$ and do another scaling with a parameter $\lambda_2\neq\lambda_1$. It can be easily seen that by multiplying $\mathbf{A}_1(\lambda_0=1,\lambda_1)$ by $\frac{\lambda_1}{\lambda_1-1}$ and $\mathbf{A}_1(\lambda_0=1,\lambda_2)$ by $\frac{\lambda_2}{\lambda_2-1}$ the term proportional to $\hbar^7$ can be vanished:
\begin{equation}
\begin{aligned}
    &\mathbf{A}_2(\lambda_0=1,\lambda_1,\lambda_2):=\frac{\lambda_1}{\lambda_1-1}\mathbf{A}_{1}(\lambda_0=1,\lambda_1)-\frac{\lambda_2}{\lambda_2-1}\mathbf{A}_{1}(\lambda_0=1,\lambda_2)=\\
    &=\frac{\lambda_2-\lambda_1}{(\lambda_1-1)(\lambda_2-1)}\mathcal{A}^\mathcal{K}_{(\vec{\alpha}|\vec{\beta})}(\hbar)-\frac{\lambda_1}{\lambda_1-1}\mathcal{A}^\mathcal{K}_{(\lambda_1\vec{\alpha}|\lambda_1\vec{\beta})}\left(\frac{\hbar}{\lambda_1}\right)+\frac{\lambda_2}{\lambda_2-1}\mathcal{A}^\mathcal{K}_{(\lambda_2\vec{\alpha}|\lambda_2\vec{\beta})}\left(\frac{\hbar}{\lambda_2}\right)=O(\hbar^6)\,.
\end{aligned}
\end{equation}
Proceeding this way, we get the Alexander scaling relation for an arbitrary representation. Write down one more step explicitly. On the third step, we turn to zero the term proportional to $\hbar^6$:
\begin{equation}
\begin{aligned}
    &\mathbf{A}_3(\lambda_0=1,\lambda_1,\lambda_2,\lambda_3):=\frac{\lambda_1\lambda_2}{\lambda_2-\lambda_1}\mathbf{A}_2(\lambda_0=1,\lambda_1,\lambda_2)-\frac{\lambda_1\lambda_3}{\lambda_3-\lambda_1}\mathbf{A}_2(\lambda_0=1,\lambda_1,\lambda_3)=\\
    &=\frac{\lambda_1(\lambda_3-\lambda_2)}{(\lambda_1-1)(\lambda_2-1)(\lambda_3-1)}\mathcal{A}^\mathcal{K}_{(\vec{\alpha}|\vec{\beta})}(\hbar)-\frac{\lambda_1^3(\lambda_3-\lambda_2)}{(\lambda_1-1)(\lambda_2-\lambda_1)(\lambda_3-\lambda_1)}\mathcal{A}^\mathcal{K}_{(\lambda_1\vec{\alpha}|\lambda_1\vec{\beta})}\left(\frac{\hbar}{\lambda_1}\right)+\\
    &+\frac{\lambda_1\lambda_2^2}{(\lambda_2-1)(\lambda_2-\lambda_1)}\mathcal{A}^\mathcal{K}_{(\lambda_2\vec{\alpha}|\lambda_2\vec{\beta})}\left(\frac{\hbar}{\lambda_2}\right)-\frac{\lambda_1\lambda_3^2}{(\lambda_3-1)(\lambda_3-\lambda_1)}\mathcal{A}^\mathcal{K}_{(\lambda_3\vec{\alpha}|\lambda_3\vec{\beta})}\left(\frac{\hbar}{\lambda_3}\right)=O(\hbar^8)\,.
\end{aligned}
\end{equation}
And so on. One can write the explicit formula for the step $k\,$:
\begin{equation}
\begin{aligned}
    &\mathbf{A}_k(\lambda_0,\lambda_1,\dots,\lambda_k)=\frac{\lambda_{k-2}\lambda_{k-1}}{\lambda_{k-1}-\lambda_{k-2}}\mathbf{A}_{k-1}(\lambda_0,\lambda_1,\dots,\lambda_{k-2},\lambda_{k-1})-\frac{\lambda_{k-2}\lambda_{k}}{\lambda_{k}-\lambda_{k-2}}\mathbf{A}_{k-1}(\lambda_0,\lambda_1,\dots,\lambda_{k-2},\lambda_k)=\\&=\sum\limits_{i=0}^{k-2}\frac{\lambda_i^k(\lambda_k-\lambda_{k-1})\prod\limits_{j\neq i}^{k-2}\lambda_j}{\prod\limits_{j\neq i}^k(\lambda_j-\lambda_i)}\mathcal{A}^\mathcal{K}_{(\lambda_i\vec{\alpha}|\lambda_i\vec{\beta})}\left(\frac{\hbar}{\lambda_i}\right)+\\
    &+(-1)^k\left\{\frac{\lambda_k^{k-1}\prod\limits_{j=0}^{k-2}\lambda_j}{\prod\limits_{j=0}^{k-2}(\lambda_k-\lambda_j)}\mathcal{A}^\mathcal{K}_{(\lambda_k\vec{\alpha}|\lambda_k\vec{\beta})}\left(\frac{\hbar}{\lambda_k}\right)-\frac{\lambda_{k-1}^{k-1}\prod\limits_{j=0}^{k-2}\lambda_j}{\prod\limits_{j=0}^{k-2}(\lambda_{k-1}-\lambda_j)}\mathcal{A}^\mathcal{K}_{(\lambda_{k-1}\vec{\alpha}|\lambda_{k-1}\vec{\beta})}\left(\frac{\hbar}{\lambda_{k-1}}\right)\right\}=O(\hbar^{k+4}).
\end{aligned}
\end{equation}
And in the limit $k\rightarrow\infty$ we get the relation, considering $\lambda_j-\lambda_{j-1}=1$:
\begin{equation}
    \sum\limits_{i=0}^{\infty}\prod\limits_{j\neq i}^{\infty}\frac{\lambda_i}{(\lambda_j-\lambda_i)}\mathcal{A}^\mathcal{K}_{(\lambda_i\vec{\alpha}|\lambda_i\vec{\beta})}\left(\frac{\hbar}{\lambda_i}\right)=0\,.
\end{equation}





\setcounter{equation}{0}
\section{Conclusion/Discussion}\label{discussion}
In our previous article~\cite{LST1}, we have studied in detail the weight system $\varphi_{\mathfrak{sl}_N}^R(\mathcal{D})$ using the constraints coming from the colored HOMFLY symmetries. In this paper lots of consequences for quantum knot invariants have been obtained. Below we list a number of concrete results of this work:
\begin{itemize}
    \item We have found the HOMFLY group factors in the basis associated with the Jacobi diagrams for an arbitrary representation $R$ up to the 6-th order (see Subsection~\ref{GF-Jacobi}).
    \item We have presented the unified combinatorial algorithm for the HOMFLY group factors' construction (see Subsection~\ref{CombConstr}). It has clear dependence on the rank $N$ and the representation $R$.
    \item As, using our method, one can find the group factors of the HOMFLY polynomials up to higher levels, we have presented an algorithm for computing Vassiliev invariants up to higher orders (see Section~\ref{MoreVI}). We have demonstrated this algorithm by computing Vassiliev invariants up to the 11-th order for the knot $3_1$ and up to the 10-th order for the knot $5_2$~\cite{knotebook}. However, beginning with the 8-th level we are not able to write down all Vassiliev invariants due to the coincidence of the primary group factors. 
    \item We have discussed the manifestation of Vogel's theorem of not distinguishing Vassiliev invariants by semisimple Lie (super)algebras' weight systems with the use of concrete examples for the $\mathfrak{sl}_N$ weight system (see Section~\ref{Vogel}).
    \item We have introduced the method of finding recursive relations for the Jones polynomials and applied it for the special case of the torus knots $T[2,2k+1]$ (see Section~\ref{Jones}).
    \item We have noticed the existence of another symmetry of the Alexander polynomials coming from restrictions applied by the conjugation symmetry of the HOMFLY polynomials (see Subsection~\ref{NewSymmAlexander}).
    \item We have studied $N$-deformation from the Alexander polynomial to the HOMFLY polynomial (see Subsection~\ref{AlToHOMFLY}).
    \item We have come up with a generalization of the currently known 1-hook scaling property for the Alexander polynomials (see Subsection~\ref{AlScaling}).
\end{itemize}
The aforementioned results present new questions for our future research. Here are some of them:
\begin{itemize}
    \item Find the novel Alexander symmetry described in Subsection~\ref{NewSymmAlexander} which forbids $Y_{[1]}^{2k+1}$.
    \item Construct a basis in the space of chord diagrams up to higher levels (at least up to the 8-th level) and find out whether Vogel's parametrization solves the problem of not distinguishing the primary Vassiliev invariants by simple Lie algebras' weight systems~\cite{VOGEL20111292}.
    \item Invent a method for constructing recursive relations for the colored HOMFLY polynomials. One can start with the symmetric and $[r,r]$ representation and the torus knots $T[2,2k+1]$. The fact that the HOMFLY polynomials are $q$-holonomic functions was proved in~\cite{garoufalidis2018colored}.
    \item Construct a recursive relation for the Jones polynomials for $p$-twisted knots in general form. Note that for several fixed values of $p$ recursive relation for the Jones polynomials have already been computed~\cite{garoufalidis2010non}, but there is still no general analytical formula or a combinatorial construction for an arbitrary value of $p$.  
    \item Verify implementation of the Plucker relations for the HOMFLY group structure.
    \item Establish a connection between group factors of the HOMFLY polynomials and integrable systems. One can look for a concrete example of connection between the Alexander polynomials and the KP hierarchy in~\cite{Mishnyakov:2019otq}.
\end{itemize}

\section{Acknowledgments}
We would like to thank A.Yu. Morozov, V.V. Mishnyakov and A.V. Popolitov for useful discussions and interesting questions which motivate us for future research. We also thank E.M. Bazanova for English language editing of this manuscript. This work was funded by the Russian Science Foundation (Grant No.20-71-10073).

\printbibliography

\end{document}